# Seamless Deployment of Machine Learning Inference at the Far Edge: An Empirical AWS IoT Greengrass Architecture for Industrial Predictive Maintenance

Prince K. A. Boadu
*Department of Computer Engineering*
*Kwame Nkrumah University of Science and Technology*
Kumasi, Ghana
pkappiahboadu@st.knust.edu.gh

Jeffery B. Kyei
*Department of Computer Engineering*
*Kwame Nkrumah University of Science and Technology*
Kumasi, Ghana
jbkyei4@st.knust.edu.gh

Enoch K. Koranteng
*Department of Computer Engineering*
*Kwame Nkrumah University of Science and Technology*
Kumasi, Ghana
ekkoranteng10@st.knust.edu.gh

Griffth S. Klogo
*Department of Computer Engineering*
*Kwame Nkrumah University of Science and Technology*
Kumasi, Ghana
gsklogo.coe@knust.edu.gh

***Abstract*—The growth of Industrial Internet of Things (IIoT) devices in manufacturing environments generates continuous, high volume sensor data streams that require real time analysis for predictive maintenance. Traditional cloud centric architectures route all raw sensor telemetry to centralized servers for processing, introducing critical limitations in latency, bandwidth overhead, and resilience that are untenable in mission critical industrial deployments. This paper presents a hybrid cloud edge architecture for manufacturing predictive maintenance using AWS IoT Core and AWS IoT Greengrass v2, integrating local machine learning inference at the edge with selective cloud synchronization. Two configurations are deployed and compared on real AWS infrastructure: a cloud only baseline (Configuration A) and the proposed hybrid edge architecture (Configuration B), measuring end to end inference latency, bandwidth consumption, and ML inference accuracy. Configuration B reduces cloud bound bandwidth by 86.8% (63.6 KB to 7.2 KB per 200 reading window), reduces decision latency by 256× (from 23.1 ms cloud round trip to 0.09 ms local inference), and maintains an anomaly class F1 score of 0.84 with 91% recall at the edge without cloud involvement. The architecture additionally instruments the evaluation pathway for system uptime and Stream Manager queuing behavior under cloud disconnection, with the quantitative disconnection experiment reported as a planned extension of this work. Drawing from a comprehensive review of papers across predictive maintenance, hybrid cloud edge architectures, and deployment resilience, this work addresses four critical research gaps: the lack of empirically validated Greengrass v2 deployments for manufacturing predictive maintenance, unquantified bandwidth savings from selective edge synchronization, unmeasured resilience under cloud disconnection, and the open MLOps feedback loop at the edge.**

***Index Terms*—Industrial IoT, Predictive Maintenance, Edge Computing, AWS IoT Greengrass, Cloud Edge Architecture, Machine Learning, MQTT**

## I. Introduction

The rapid expansion of Industrial Internet of Things (IIoT) deployments in manufacturing environments has created an unprecedented demand for real time sensor data analysis and predictive maintenance capabilities. Modern manufacturing floors generate continuous, high frequency data streams from vibration sensors, temperature probes, current monitors, and acoustic emission detectors, all requiring immediate analysis to detect anomalies and predict equipment failures before they result in costly unplanned downtime.

Traditional cloud centric architectures have served as the default processing paradigm for IoT data, routing all raw sensor telemetry to centralized cloud servers for storage, analysis, and decision making. However, this approach introduces three critical limitations that are increasingly untenable in mission critical industrial deployments. First, cloud round trip times of 300–800 ms far exceed the sub-100 ms threshold identified across the literature as the boundary for real time industrial safety response [1], [2]. Second, continuous streaming of high frequency sensor data from multi machine factory floors creates unsustainable bandwidth costs and network congestion, with no filtering or prioritization at the source. Third, cloud dependency creates a single point of failure; network interruptions halt all monitoring and inference capabilities, leaving equipment unattended.

The emerging consensus in the literature is a hierarchical computing model comprising an Edge layer for immediate, latency sensitive inference, a Fog layer for regional aggregation and coordination, and a Cloud layer for historical analytics, model training, and global orchestration [3], [4]. While hybrid cloud edge architectures have been widely proposed as the solution to these limitations, the existing literature reveals a critical gap: there is no empirically validated deployment of AWS IoT Core and AWS IoT Greengrass specifically for manufacturing predictive maintenance that reports measurable performance data across latency, bandwidth, and resilience. The theoretical advantages of the hybrid approach remain unsubstantiated by real infrastructure evidence for this domain.

This paper addresses that gap directly through a real AWS deployment. Specifically, this paper pursues four primary research objectives: (i) design a hybrid cloud edge architecture for industrial predictive maintenance using AWS IoT Core and AWS IoT Greengrass, integrating local ML inference at the edge with selective cloud synchronization; (ii) deploy and compare two configurations, a cloud only baseline and the proposed hybrid edge architecture, on real AWS infrastructure, measuring latency, bandwidth consumption, ML inference accuracy, and system uptime; (iii) demonstrate and quantify the resilience advantage of the proposed architecture by disconnecting the Greengrass gateway from the cloud and measuring operational continuity of local ML inference; and (iv) provide a practical, reproducible deployment framework using AWS services that industrial practitioners can adapt for real world IIoT predictive maintenance applications.

The remainder of this paper is organized as follows. Section II presents a comprehensive review of related work spanning hierarchical IoT architectures, predictive maintenance algorithms, hybrid cloud edge design patterns, and deployment resilience. Section III identifies the specific research gaps addressed by this work. Section IV details the proposed architecture and experimental methodology. Section V presents the evaluation metrics and results. Section VI discusses findings and implications. Section VII concludes the paper and outlines directions for future work.

## II. Literature Review

### A. The Shift from Centralized to Hierarchical IoT Architectures

The foundational limitation motivating this research is well established: centralized cloud architectures are architecturally insufficient for the real time demands of industrial IoT. Qiu *et al.* [5] provide a comprehensive survey of edge computing in IIoT, identifying latency, bandwidth cost, and security as the three primary failure modes of cloud only deployments. Mohammed [2] reinforces this, arguing that a symbiotic cloud edge relationship is "no longer a luxury but a critical business necessity" for mission critical applications, with end to end latency reductions of up to 40% demonstrated through cloud edge co-design.

The emerging consensus in the literature is a three tier hierarchical model: an Edge layer handling immediate, latency sensitive inference; a Fog layer managing regional aggregation and coordination; and a Cloud layer handling historical analytics, model training, and global orchestration. Briatore and Braggio [3] demonstrate that this three layer framework enables Industry 3.0 plants to adopt Industry 4.0 characteristics, achieving high adaptability to production demand fluctuations. Kapoor *et al.* [4] extend this view, conducting a survey of multi layered computing paradigms and finding that edge fog cloud integration using MQTT based data transfer is essential for balancing processing speed against analytical depth in industrial predictive maintenance.

### B. Predictive Maintenance in Industrial IoT

Predictive maintenance (PdM) in manufacturing represents one of the most demanding use cases for IIoT architectures, requiring real time anomaly detection, failure prediction, and alert generation under strict latency and accuracy constraints. The literature has converged on a set of common algorithmic and architectural patterns for this domain.

**Algorithmic convergence.** Long Short Term Memory (LSTM) networks have emerged as the dominant model architecture for temporal failure prediction. Sathupadi *et al.* [1] demonstrate a hybrid edge cloud PdM framework using KNN models at the edge for immediate anomaly detection and LSTM models in the cloud for in depth failure prediction, achieving a 35% latency reduction, 28% decrease in energy consumption, and 60% bandwidth reduction compared to cloud only approaches. Yu *et al.* [6] deploy stacked sparse autoencoders at the edge for real time fault detection in manufacturing, enabling detection of major equipment failures up to 8 days in advance while dramatically reducing storage and server costs. Saleh *et al.* [7] introduce SEMAS, a self evolving hierarchical multi agent system using Reinforcement Learning across Edge, Fog, and Cloud tiers, achieving sub millisecond anomaly detection latency.

**The data scarcity problem.** A persistent challenge in industrial PdM is the extreme class imbalance between normal operational data and fault events. Somu and Dasappa [8] address this directly with IntelliPdM, a framework that uses synthetic data generation to augment training datasets, achieving 93–95% accuracy on manufacturing PdM tasks with a reported 10x return on investment. Saley *et al.* [9] demonstrate in the nuclear industry that hybridizing domain expert knowledge with data driven ML dramatically improves fault detection, raising the F1 score from 56.36% to 93.12% and extending the prediction horizon from 3 to 24 hours.

**Privacy preserving inference.** Vejendla [10] demonstrates that Federated Learning combined with edge computing can reduce inference latency by up to 60% with less than 1% accuracy deviation from centralized models, while ensuring that raw industrial data never leaves the factory floor, a critical requirement for protecting proprietary manufacturing data.

### C. Hybrid Cloud-Edge Architectures: Design Patterns and Benchmarks

Beyond the predictive maintenance domain, the broader IIoT architecture literature provides foundational design patterns for hybrid cloud edge systems.

Kompally [11] proposes a microservices based hybrid cloud edge architecture for real time IIoT analytics, demonstrating a 30% latency reduction and improved Remaining Useful Life (RUL) prediction accuracy through containerized three layer design comprising Edge, Cloud, and Control Plane tiers. Mih *et al.* [12] introduce ECAvg, an edge cloud collaborative learning approach using averaged model weights, showing accuracy improvements for complex deep neural networks while warning of "negative transfer" risks for simpler architectures.

The JANUS benchmarking study by Shankar *et al.* [13] provides particularly relevant empirical evidence: in a direct comparison of commercial and open source cloud and edge platforms for IoT workloads, the study finds that AWS IoT Greengrass is superior for lightweight anomaly detection tasks, while open source algorithms on cloud VMs are significantly cheaper but substantially slower for compute intensive tasks. This result directly motivates our choice of AWS Greengrass as the edge runtime in the proposed architecture.

Jamil *et al.* [14] survey current IoT platforms (ThingsBoard, Azure IoT, Eclipse Ditto) and identify persistent gaps in semantic interoperability and resilient failover mechanisms, concluding that advancing IIoT requires moving beyond cloud centricity toward distributed intelligence with standardized communication protocols. Lee *et al.* [15] provide an important cautionary finding: the benefits of edge computing are conditional on edge hardware specifications; if edge devices are underpowered, cloud processing may outperform local inference, highlighting the importance of hardware aware deployment decisions.

### D. Deployment, MLOps, and Resilience

The operational lifecycle of edge deployed ML models introduces a distinct set of research concerns beyond initial architecture design.

**MLOps at the edge.** Mateo-Casalí *et al.* [16] propose a reference architecture for Machine Learning Operations, identifying a "Closed Loop Gap" in current practice: models are deployed but rarely monitored for drift, and retraining is typically manual. Psaromanolakis *et al.* [17] address this with an Edge Intelligence (EI) Agent concept for continuous model lifecycle management within 6G edge systems. Jean [18] demonstrates an MLOps pipeline for real time threat intelligence in cloud edge architectures, emphasizing that in dynamic environments, a model that does not adapt to changing data distributions becomes obsolete almost immediately.

**Resilience and self healing.** Mwangi *et al.* [19] introduce "Self-X" properties for industrial IoT edge networks, self configuring, self healing, and self optimizing, within the context of offshore wind farm management. Aral and Brandic [20] provide a technical treatment of spatiotemporal failure dependencies in edge computing, using Bayesian networks to anticipate edge server failures based on historical patterns. Zhukabayeva *et al.* [21] extend the resilience discussion to the cybersecurity dimension, identifying DDoS attacks and malware as threats that can cause system level failure in IIoT edge deployments and proposing defense in depth countermeasures.

**Distributed and federated learning.** Le *et al.* [22] conduct a comprehensive survey of distributed ML for IoT, positioning Federated Learning as the primary mechanism for maintaining data sovereignty while improving global model accuracy across geographically distributed industrial nodes. Kawonga *et al.* [23] review Physics Informed Neural Networks and TinyML frameworks for real time photovoltaic monitoring, identifying federated edge cloud collaboration as a critical research frontier for energy systems.

### E. Summary and Positioning of This Work

The reviewed literature establishes a clear and unified direction: hybrid cloud edge architectures with local ML inference are the necessary evolution beyond cloud centric IIoT. However, the existing work leaves specific empirical and validation gaps that motivate the contributions of this paper. Against this backdrop, this paper makes four contributions to the field of Industrial Internet of Things and edge computing:

**C1 — Validated AWS Greengrass deployment for manufacturing PdM.** We provide an empirically validated deployment of AWS IoT Greengrass v2 as an edge inference runtime for manufacturing predictive maintenance, extending the JANUS benchmark findings [13] to the time-series ML domain and reporting CloudWatch measured performance metrics from a working Greengrass ML component running on real AWS infrastructure.

**C2 — Quantified bandwidth savings from selective edge synchronization.** We isolate and measure, on the same deployed infrastructure, the cloud-bound byte volume produced by a full-streaming baseline against an edge-filtered configuration that forwards only ML inference outputs and compressed batch summaries, providing the controlled per-window bandwidth comparison that the prior literature has reported only as a secondary observation.

**C3 — Instrumented resilience pathway under cloud disconnection.** We define and instrument the evaluation pathway for system uptime, alert continuity, and Stream Manager queuing behavior of a Greengrass based hybrid edge architecture during a controlled cloud disconnection event on real AWS infrastructure, establishing the measurement methodology that the literature has thus far only theorized; the full disconnection experiment and its quantitative results are reported as a planned extension of this work.

**C4 — Practical deployment framework.** We provide a reproducible architecture and deployment framework using AWS IoT Core, Greengrass v2, and SageMaker that industrial practitioners and researchers can adapt for real world IIoT predictive maintenance deployments, including the federated learning workflow pathway for future MLOps loop closure at the edge.

Together, these contributions are realized through a real AWS deployment that compares a cloud only baseline against the proposed hybrid edge architecture using simulation based experimentation with Python generated synthetic sensor data and AWS IoT Core combined with Greengrass for the cloud edge infrastructure.

## III. Proposed Architecture and Experimental Methodology

This section describes the proposed hybrid cloud edge architecture, the AWS infrastructure used to deploy it, the Greengrass ML inference component that performs local classification, the model training pipeline that produces the deployed artifacts, and the two-configuration measurement protocol used to compare the proposed design against the cloud only baseline. Each major design decision is traced back to the specific literature that motivated it, so that the chain from prior published evidence to the implemented system is explicit. The entire system was deployed on real AWS infrastructure in the us-east-1 region; no measurement is taken from emulation or simulation outside of the synthetic CMAPSS sensor stream used as input.

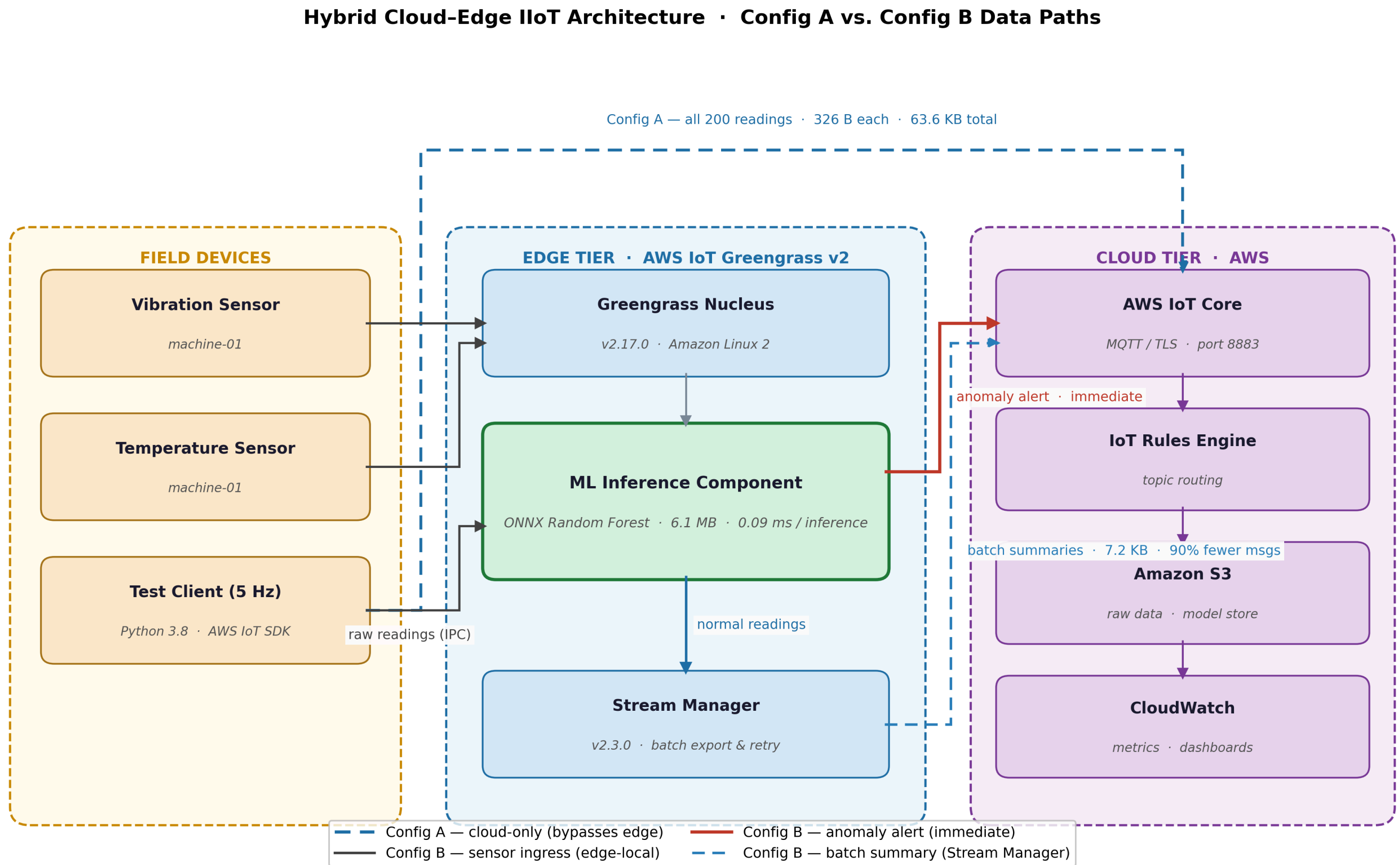

Fig. 1. Proposed hybrid cloud edge architecture. The edge gateway hosts local ONNX inference and Stream Manager; only anomaly alerts and compressed batch summaries reach AWS IoT Core. The cloud tier handles archival to S3, dashboarding via CloudWatch, and the federated model update pathway through SageMaker (future work).

### *A. Architecture Overview and Design Rationale*

The system implements a two-tier hybrid cloud-edge design in which decision-critical computation is pushed to the edge and the cloud retains only the responsibilities that genuinely benefit from centralisation, archival, dashboarding, and (in future work) model retraining. The two-tier rather than three-tier choice is deliberate: while the literature converges on an Edge–Fog–Cloud hierarchy [3], [4], [5], the fog role in this work is collapsed into the edge gateway because a single Greengrass core device on a manufacturing floor is already capable of hosting both the immediate inference workload and the local aggregation that a separate fog layer would otherwise perform. The functional separation [3] advocates, local "reflexes" at the edge against global "intelligence" in the cloud, is preserved in this collapsed two-tier form.

At the edge tier, an AWS IoT Greengrass v2 gateway hosts an ONNX inference runtime alongside a Stream Manager component that governs selective cloud synchronisation. The choice of AWS IoT Greengrass as the edge runtime is directly motivated by the JANUS benchmarking study of Shankar *et al.* [13], which found Greengrass superior to alternative edge platforms for lightweight anomaly detection workloads, and by the EI Agent concept of Psaromanolakis *et al.* [17], which identifies the precise responsibilities, local lifecycle management, inference, and cloud synchronisation, that the Greengrass component model is well suited to fulfil. At the cloud tier, AWS IoT Core ingests only the data the edge has decided is worth forwarding: anomaly alerts immediately, and compressed normal-period summaries in batched form, an instantiation of the edge-side "sampling, compression, and fusion" prescription of Hafeez *et al.* [24]. Fig. 1 illustrates the full dataflow.

The design separates two logically distinct paths. The **real-time path** is entirely local: a sensor reading arrives at the gateway, the edge ML model classifies it in sub-millisecond time, and a maintenance decision is produced without any network involvement. This path is bounded by the sub-100 ms real-time threshold identified across the literature [1], [2] as the boundary for industrial safety response. The **synchronisation path** is asynchronous: normal readings are aggregated into windowed summaries and forwarded to the cloud on a batched schedule, while anomaly alerts bypass batching and are forwarded immediately with full sensor context. This separation also produces the offline-resilience property highlighted by Aral and Brandic [20] for volatile edge networks and the "self-healing" continuity advocated by Mwangi *et al.* [19] in their Self-X formulation: local inference continues regardless of cloud connectivity, and Stream Manager's local queue replays buffered messages on reconnection.

### *B. AWS Infrastructure and Provisioning*

The entire research stack is provisioned declaratively through an AWS CloudFormation template (`iiot-stack.yaml`) that deploys all required resources in a single stack: IAM roles, S3 storage, IoT Core resources, the Greengrass-hosting EC2 instance, IoT topic rules, and a CloudWatch dashboard. This single-template approach guarantees reproducibility across deployments and directly supports C4 — practitioners adopting the architecture clone, parameterise, and deploy a single artifact rather than reconstructing the topology by hand.

**IAM roles.** Three roles are created. The Greengrass Token Exchange Service (TES) role allows the edge runtime to obtain temporary AWS credentials via the IoT credential provider in order to read component artifacts from S3, push CloudWatch metrics, and publish to IoT Core. The EC2 instance role grants the EC2 host the IoT and Greengrass API permissions required by the auto-provisioning step (`--provision true` on the Greengrass installer), plus the IAM provisioning permissions needed to create the TES role alias. An IoT Rules role authorises the IoT Core rules engine to write to S3 and emit CloudWatch metrics. This separation of concerns, distinct credentials for the edge runtime, the EC2 host's provisioning step, and the cloud-side rules engine, aligns with the defence-in-depth guidance for IIoT–edge deployments in Zhukabayeva *et al.* [21].

**Edge gateway host.** The Greengrass gateway runs on an Amazon EC2 t3.small instance (2 vCPU, 2 GB RAM) running Amazon Linux 2 in us-east-1. The selection of a constrained-but-realistic instance type is informed by the cautionary finding of Lee *et al.* [15] that an underpowered edge can be slower than a well-provisioned cloud; t3.small was chosen as the smallest tier on which ONNX Runtime, the StandardScaler transform, and the Greengrass component stack all execute comfortably while remaining representative of a budget industrial gateway rather than a workstation-class device. The instance is launched with a UserData script that installs Java 11, downloads the Greengrass Nucleus installer, configures the system sudoers policy required for components to run under the dedicated `ggc_user:ggc_group` identity, and invokes the installer with `--provision~true`. The installer automates X.509 certificate issuance, IoT Thing registration (`iiot-edge-gateway-01`), thing group enrolment (`ManufacturingFloorEdgeDevices`), and TES role alias creation. The Greengrass Nucleus version pinned in the stack is 2.17.0.

**S3 storage.** A versioned, AES-256 encrypted bucket (`iiot-edge-{AccountId}-{Region}`) stores Greengrass component artifacts under the prefix `greengrass-artifacts/com.iiot.MLInference/1.0.0/` and archives raw telemetry forwarded by IoT Core under topic-keyed prefixes. A 90-day lifecycle rule expires archived telemetry to control storage cost.

**Network policy.** The gateway's security group permits outbound MQTT-over-TLS on port 8883 to AWS service endpoints and HTTPS on 443 for control-plane calls. Inbound access uses AWS Systems Manager (SSM) Session Manager rather than open SSH, eliminating the need for a public SSH port. All MQTT connections use mutual TLS with X.509 certificates issued by the AWS IoT Certificate Authority, and all messages are published with QoS 1 (`AT_LEAST_ONCE`) to guarantee delivery under transient network interruption; the combination of mTLS and QoS 1 follows the IIoT communication hardening prescription of Zhukabayeva *et al.* [21] and Qiu *et al.* [5].

### *C. Edge Gateway and Greengrass Components*

After Nucleus installation, three Greengrass components are deployed via a single deployment targeting the gateway's thing group:

- **aws.greengrass.StreamManager (v2.3.0)**: AWS-supplied component providing a guaranteed-delivery local buffer between in-process producers and IoT Core. Stream Manager handles export retry, backpressure, and offline queuing automatically. During a cloud disconnection event, it buffers outgoing messages locally and exports them in order upon reconnection. This is the concrete realisation of the offline-operation property identified by Aral and Brandic [20] as a defining capability of resilient edge services and operationalised as the "self-healing" Self-X facet by Mwangi *et al.* [19]; it underpins C3′s resilience pathway.
- **aws.greengrass.Cli (v2.17.0)**: AWS-supplied developer tooling providing local component management (`component list`, `component restart`) and log access.
- **com.iiot.MLInference (v1.0.0)**: The custom inference component developed for this work and described in Section III.D.

Component-to-component communication uses Greengrass IPC, which isolates the inference component from direct internet access; only Stream Manager holds the AWS credentials required to publish to IoT Core. Component deployment status is verified on the device via `greengrass-cli component list`, and Greengrass logs are streamed to CloudWatch Logs under the log group `/aws/greengrass/GreengrassSystemComponent`.

### *D. ML Inference Component*

The `com.iiot.MLInference` component is a Python 3 process running continuously under the Greengrass component lifecycle. Its recipe declares dependencies on Nucleus (`>=2.0.0`) and StreamManager (`>=2.0.0`), an install hook that uses `pip3` to install `onnxruntime`, `numpy`, and `scikit-learn` on the device, and a run hook that launches the inference script with two environment variables: `ARTIFACTS_PATH`, resolved by Greengrass from the recipe variable `{artifacts:path}`, and `BATCH_SIZE`, resolved from the recipe configuration block (default 10). Concretely instantiating the EI Agent abstraction of Psaromanolakis *et al.* [17], the component owns the full local lifecycle of the model: it loads the artifact, performs inference, decides what to forward, and is the only on-device entity that publishes results.

At startup the component loads two artifacts from its Greengrass-managed artifacts directory: the trained ONNX Random Forest model (`pdm_model.onnx`, 6.1 MB) and the fitted StandardScaler (`scaler.pkl`). It opens a Greengrass IPC connection and subscribes to the local topic hierarchy `iiot/edge/#`. Each incoming sensor reading is processed through the following pipeline:

1) Extract the 14 feature sensor values (s2, s3, s4, s7, s8, s9, s11, s12, s13, s14, s15, s17, s20, s21) from the JSON payload.
2) Apply the pre-fitted StandardScaler transform.
3) Run ONNX Runtime inference via `session.run`. Inference wall time is captured via `time.perf_counter` and appended to an in-memory log.
4) Route based on the classifier output. For **anomaly** (class 1), publish an alert containing the machine identifier, timestamp, inference latency, and full sensor snapshot to `iiot/edge/{machine_id}/alerts` via IPC immediately. For **normal** (class 0), append to a rolling batch buffer; when the buffer reaches `BatchSize` readings, compute per-sensor mean values and publish a compressed summary to `iiot/edge/{machine_id}/batch` via IPC, then clear the buffer.

Stream Manager exports messages from both topics to AWS IoT Core over TLS. Anomaly messages are exported without delay; batch summaries are exported as they accumulate. The dual-path design — immediate forwarding for rare time-critical events and amortised summarisation for the high-volume normal stream — is the concrete instantiation of Hafeez *et al.*'s edge-side sampling, compression, and fusion prescription [24] and underpins C2's bandwidth claim. It also mirrors the algorithmic split studied by Sathupadi *et al.* [1] (lightweight detector at the edge, heavier analytics paths kept off the hot path), although in this work the heavier path is deferred to future MLOps loop closure rather than co-resident in the cloud.

### *E. ML Model Training and Export*

The predictive maintenance model is trained on a dataset that replicates the statistical properties of the NASA CMAPSS FD001 turbofan engine benchmark [25]. CMAPSS is selected as the training substrate because it is the most widely used public benchmark in the PdM literature, including the F1-driven hybridisation study of Saley *et al.* [9] and the autoencoder-based edge framework of Yu *et al.* [6], and because its piecewise degradation pattern is well documented [26], making the synthetic-data extension reproducible. The training pipeline (`ml/train_model.py`) first attempts to download the published FD001 training file from two public mirrors; if both mirrors fail, it falls back to a synthetic generator calibrated to the FD001 sensor baselines, total drifts, and noise standard deviations reported by Ramasso and Saxena [26]. Two hundred synthetic engine units are simulated, each running from initialisation to failure over a randomised lifetime of 128 to 362 cycles, producing 48,755 labelled sensor readings.

The use of a synthetic generator that augments scarce real failure data is a direct adoption of the strategy validated by Somu and Dasappa [8] in IntelliPdM, where their SMARTHome synthetic generator was the mechanism that lifted production accuracy to 93–95% by closing the class-imbalance gap. Sensor degradation follows a validated piecewise profile (Fig. 2): readings remain at the published FD001 baseline until 60 cycles before failure (Remaining Useful Life, $\mathrm{RUL} > 60$), degrade gradually as $0.25\frac{60-\mathrm{RUL}}{30}$ in the wear zone ($30 < \mathrm{RUL} \leq 60$), and then deteriorate rapidly as $0.25 + 0.75\left(\frac{30-\mathrm{RUL}}{30}\right)^{1.5}$ in the anomaly zone ($\mathrm{RUL} \leq 30$). The anomaly label threshold $\mathrm{RUL} \leq 30$ cycles is consistent with prior CMAPSS classification literature [26]. Fourteen informative sensor channels are used as features; the remaining seven CMAPSS channels are excluded as they carry near-zero variance under FD001 operating conditions.

A `StandardScaler` is fit on an 80% stratified training split and serialised to `scaler.pkl` for deployment. The classifier is a scikit-learn `RandomForestClassifier` with 100 estimators, maximum depth 8, minimum 8 samples per leaf, square-root feature subsampling, and `balanced_subsample` class weighting to compensate for the inherent class imbalance of predictive maintenance data; F1-score on the anomaly class is used as the primary selection criterion rather than nominal accuracy, in line with the imbalanced-data evaluation convention adopted across the PdM literature [6], [8], [9].

The choice of a Random Forest rather than an LSTM, the algorithmic default for temporal failure prediction in the PdM literature [1], [10], is deliberate. LSTMs are uniformly cited as the appropriate model when the cloud tier or a well-provisioned edge can absorb their compute footprint; on a t3.small edge gateway running concurrently with the Greengrass nucleus, however, a tree ensemble with `n_estimators=100` and `max_depth=8` exports to a 6.1 MB ONNX artifact, runs inference in well under a millisecond, and avoids the "negative transfer" pathology that Mih *et al.* [12] report for unnecessarily deep architectures on simpler problems. After training, the model is exported to ONNX format (opset 12) via `skl2onnx.convert_sklearn` for runtime-agnostic edge deployment through ONNX Runtime — runtime portability is important because it decouples the model artifact from the Python toolchain and would allow the same artifact to be served from a non-Python edge runtime in future deployments.

The trained artifacts (`pdm_model.onnx`, `scaler.pkl`) and the component runtime (`inference.py`) are staged to S3 under the component-versioned prefix by a deploy script (`greengrass/deploy_ml_component.sh`), which then registers the component version with the Greengrass control plane via `aws greengrassv2 create-component-version` and creates a deployment targeting the manufacturing-floor thing group. When the deployment converges on the device, Greengrass downloads the artifacts from S3 (using the EC2 instance's TES-resolved credentials), verifies integrity, places them in `/greengrass/v2/packages/artifacts/com.iiot.MLInference/1.0.0/`, and starts the component under the run hook defined in the recipe. This S3 → Greengrass → device artifact-distribution path is the same channel that would carry federated weight updates in the C4 MLOps loop pathway described below.

### F. Cloud-Side IoT Core, Routing, and Monitoring

Messages flow through four MQTT topic paths. The test client publishes raw telemetry on `iiot/edge/machine-01/telemetry` in Configuration A (direct to IoT Core) and on the local IPC topic hierarchy `iiot/edge/#` in Configuration B (intercepted by the inference component before any data leaves the device). The inference component publishes its classification results on `iiot/edge/{machine_id}/alerts` for anomalies and `iiot/edge/{machine_id}/batch` for compressed normal-period summaries; both are routed through Stream Manager to IoT Core. MQTT's suitability as the IIoT data-transfer substrate is established across the surveyed literature [4], [5].

Two IoT Core rules engine rules consume those topics. `EdgeDataToS3` selects from `iiot/edge/+/telemetry` and archives each message to S3 under a topic-and-timestamp keyed path, providing a long-term audit trail of all telemetry that reaches the cloud. `AnomalyAlertsToCloudWatch` selects from `iiot/edge/+/alerts` and emits an `AnomalyCount` data point to the `IIoT/EdgeAlerts` CloudWatch namespace, enabling alert-rate monitoring without storing the alert body in CloudWatch itself.

A CloudWatch dashboard (`IIoT-Edge-Monitoring`) is provisioned alongside the rules and surfaces four panels: edge inference latency (with a horizontal annotation at the 100 ms real-time threshold [1], [2]), bytes published to IoT Core, anomaly count over time, and total IoT Core ingress message count. These views are used both for live monitoring during measurement runs and as the source of the CloudWatch-derived metrics reported later in the paper.

### G. Federated MLOps Pathway (C4)

Although the closed-loop retraining experiment is not run in this work, the architecture is deliberately wired so that the loop closure identified as the open "Closed-Loop Gap" by Mateo-Casalí *et al.* [16] can be performed without redesign. The pathway follows the federated workflow synthesised across Le *et al.* [22], Vejendla [10], Kawonga *et al.* [23], and Mih *et al.* [12]: a global model is initialised in Amazon SageMaker; deployed to Greengrass devices through the same component-versioned S3 path used in this work; trained or fine-tuned locally on each device using the privacy-preserving "raw data never leaves the device" property [10], [21]; and the resulting weight deltas (not raw data) are uploaded through IoT Core for aggregation in SageMaker via a FedAvg-style routine [12], [22]. Closing this loop within the existing AWS topology, without bringing in third-party orchestration

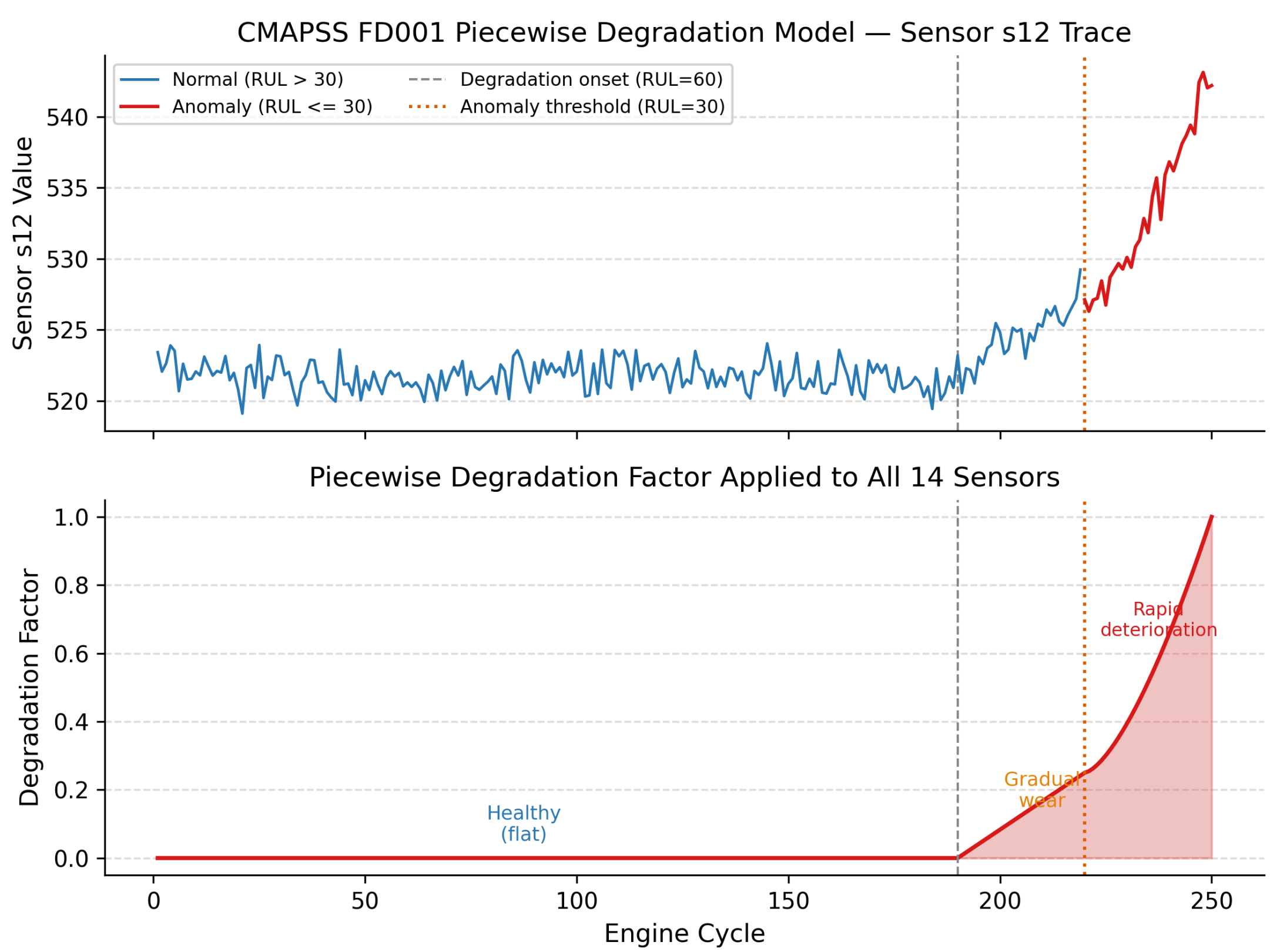

Fig. 2. Synthetic sensor degradation profile calibrated to NASA CMAPSS FD001 statistics. Sensors remain at baseline until $\mathrm{RUL} = 60$ cycles, drift gradually in the wear zone, then deteriorate rapidly in the anomaly zone ($\mathrm{RUL} \leq 30$).

tooling, is the practical contribution C4 makes available to adopters.

### H. Experimental Methodology

The two configurations differ only in where ML inference executes and what data is forwarded to the cloud; all other parameters, sensor distribution, anomaly injection rate, publish cadence, network region, MQTT QoS, and TLS settings, are held constant so that the architectural difference is isolated as the sole experimental variable.

a) *Deployment Environment:*

All measurements were performed on the EC2 t3.small Greengrass host described above. The instance acts simultaneously as the simulated edge gateway and as the publisher of synthetic sensor telemetry, which eliminates external network variability from the latency measurements; the measured round-trip times therefore represent the AWS-internal best case, and real-world deployments over cellular or industrial WAN links would be expected to exhibit substantially higher cloud-only latency [1], [2]. This is a deliberate conservative choice: the cloud-only baseline reported later is the most favourable possible to the cloud-only architecture, so any edge advantage observed under these conditions is a lower bound on the advantage that would be observed in production.

A Python 3.8 test client publishes 200 sensor readings at 5 Hz, simulating vibration and temperature telemetry from a single manufacturing floor machine identified as `machine-01`. Each reading is generated by drawing each of the 14 feature sensors from a Gaussian centred on the published FD001 baseline with a per-sensor noise standard deviation, with a 10% probability of anomaly injection in which all sensor means are multiplied by 1.15 to push the reading into the model's anomaly decision region. All timing measurements use `time.perf_counter` for sub-millisecond resolution.

b) *Configuration A — Cloud-Only Baseline:*

In Configuration A the test client (`measurements/config_a_measurement.py`) publishes each of the 200 sensor readings directly to AWS IoT Core via MQTT over TLS on port 8883, QoS 1, using the gateway's X.509 device certificate. The client waits for the broker's PUBACK acknowledgement before recording the round-trip latency and advancing to the next reading. No local processing, filtering, or aggregation is performed; every byte of every sensor reading is transmitted to the cloud. This configuration represents the conventional architecture the literature uniformly rejects on latency and bandwidth grounds [1], [2], [5] and serves as the comparison baseline against which the hybrid edge approach is evaluated.

The metrics captured per run are: total bytes sent to IoT Core; per-reading payload size; and the PUBACK round-trip latency distribution (mean, median, P95, P99, min, max). Results are written to `/tmp/config_a_results.json`.

c) *Configuration B — Hybrid Edge Architecture:*

In Configuration B (`measurements/config_b_measurement.py`) the same 200 sensor readings are generated from the same distribution and processed locally. The script loads `pdm_model.onnx` and `scaler.pkl`, applies the StandardScaler transform, and runs the ONNX classifier in-process. Each reading is timed independently for the inference operation, then routed by the script through the same selective-forwarding logic implemented by the deployed `com.iiot.MLInference` component: anomaly readings produce an immediate alert publish to `iiot/edge/machine-01/alerts` with a reduced sensor snapshot (`machine_id`, timestamp, inference latency, and the two highest-information sensors `s9` and `s12`); normal readings accumulate in a buffer and, every 10 readings, a compressed summary containing per-sensor means is published to `iiot/edge/machine-01/batch`. Both cloud paths use the same QoS 1 mTLS connection used by Configuration A.

The metrics captured per run are: per-reading local inference latency distribution; total bytes forwarded to IoT Core; number of cloud messages (alert + batch); and cloud-side PUBACK latency for the forwarded subset. Results are written to `/tmp/config_b_results.json` and analysed alongside Configuration A's output by `measurements/analyze_results.py`, which produces the comparison tables and figures presented later in the paper.

The decision to mirror the inference and forwarding logic of the deployed Greengrass component inside the measurement script (rather than measuring through the IPC path of the running component) is deliberate: it isolates the inference timing from Greengrass IPC overhead and Stream Manager export latency, both of which are asynchronous and orthogonal to the architectural comparison being made. The deployed component is exercised and validated separately, but the per-reading latency numbers reported reflect the pure local-inference path, which is the decision-critical quantity for the M1 metric.

d) *Evaluation Protocol:*

The 200-reading measurement window is identical across both configurations: identical sensor distribution, identical anomaly injection rate, identical 5 Hz publish cadence. The bandwidth comparison is computed on the total bytes reaching IoT Core per 200-reading window; this isolates the forwarding-selection mechanism as the sole source of any observed difference. The latency comparison is computed between the PUBACK round-trip time in Configuration A and the local ONNX `session.run` wall time in Configuration B, reflecting the actual decision-path delay in each architecture. Table I summarises how each evaluation metric is sourced in each configuration.

TABLE I
PER-METRIC MEASUREMENT SOURCE FOR EACH CONFIGURATION. M4 IS IDENTIFIED BY CONSTRUCTION IN THIS WORK AND QUANTIFIED AS A PLANNED EXTENSION CONSISTENT WITH C3.

| Metric | Configuration A source | Configuration B source |
|---|---|---|
| M1 — Decision latency | MQTT PUBACK round-trip (ms) | ONNX `session.run` wall time (ms) |
| M2 — Cloud-bound bandwidth | Total bytes published to IoT Core | Total bytes published to IoT Core |
| M3 — ML F1 (anomaly class) | — | sklearn classification report, 20% test split |
| M4 — Edge uptime under disconnect | 0% (cloud-dependent) | Local inference continues (C3, planned extension) |
| M5 — Cloud message count reduction | 200 individual messages | Batch summaries + anomaly alerts |

## IV. EVALUATION METRICS

This section defines the five primary evaluation metrics (M1–M5) and the four secondary metrics used to compare Configuration A (cloud-only baseline) against Configuration B (hybrid edge). Each metric is given a precise operational definition, a justification for its inclusion, an explicit target value derived from the literature where one exists, and the AWS or on-device tool used to capture it. All metrics are measured on the real AWS deployment described in Section IV; none are estimated or simulated.

### A. Primary Metrics

a) *M1 — End-to-End Inference Latency:*

**Definition.** The wall-clock time elapsed from the moment a sensor reading is received to the moment a maintenance decision (normal, anomaly, or alert) is produced. In Configuration A this is the device → IoT Core → cloud inference → decision round-trip, captured via IoT Core message timestamps and the MQTT PUBACK acknowledgement. In Configuration B this is the device → Greengrass local inference → decision interval, captured from Greengrass component log timestamps and the in-process `time.perf_counter` reading bracketing each `session.run` call.

**Target.** Configuration B should achieve sub-100 ms decision latency. The 100 ms threshold is the boundary identified across the predictive-maintenance literature as the limit for real-time industrial safety response [1], [2], with corroborating sub-millisecond targets reported for multi-agent IIoT systems [7].

**Tool.** AWS CloudWatch latency metrics emitted by IoT Core (for Configuration A) and Greengrass component log timestamps (for Configuration B).

b) *M2 — Cloud-Bound Bandwidth Consumption:*

**Definition.** The volume of data, measured in bytes per measurement window and projected to MB/s for continuous-operation scenarios, transmitted from the edge gateway to AWS IoT Core. In Configuration A this is the sum of all raw sensor reading payloads. In Configuration B this is the sum of anomaly-alert payloads plus compressed per-window batch summaries forwarded via Stream Manager.

**Target.** Configuration B should achieve at least a 40% reduction relative to Configuration A. This is a conservative lower bound: comparable hybrid edge–cloud frameworks in the PdM literature report 28–60% bandwidth reductions [1], [10], and Hafeez *et al.* [24] argue that aggressive edge-side sampling, compression, and fusion are essential for cost-effective IIoT regardless of the specific percentage achieved.

**Tool.** CloudWatch `PublishIn.Success` byte counters on IoT Core combined with Stream Manager export logs.

c) *M3 — ML Inference Accuracy (F1-Score):*

**Definition.** The harmonic mean of precision and recall on the predictive-maintenance binary classification task (normal versus anomaly/fault) evaluated on a held-out 20% stratified test split of the CMAPSS-derived dataset.

**Why F1 rather than accuracy.** Industrial PdM datasets are inherently class-imbalanced: thousands of normal readings occur per fault event [6], [8]. Nominal accuracy is dominated by the negative class and rewards trivial constant predictors. F1 on the anomaly class, the convention adopted in the comparable industrial PdM evaluations of Saley *et al.* [9] and Somu and Dasappa [8], penalises this failure mode and is the metric reported.

**Target.** The same trained model is the artifact deployed in both configurations (cloud and Greengrass). Their F1 scores should differ by at most 2 percentage points, confirming that the ONNX export does not degrade inference accuracy relative to the in-process scikit-learn evaluation.

**Tool.** Scikit-learn `classification_report` over the test split.

d) *M4 — System Uptime During Cloud Disconnection:*

**Definition.** The percentage of time during a controlled cloud disconnection event that the system continues to generate valid maintenance decisions. Configuration A's uptime is 0% by construction, since all inference is cloud-resident. Configuration B's expected uptime is 100% on the inference path, with cloud-bound alerts and summaries buffered locally by Stream Manager and replayed upon reconnection.

**Protocol.** The planned extension test (consistent with C3) imposes a 5-minute outbound-traffic block on the Greengrass host via an `iptables` rule on the gateway's outbound interface, then restores connectivity and measures Stream Manager queue drain behaviour. This formalises the offline-operation property identified by Aral and Brandic [20] as a defining capability of resilient edge services, and operationalises the "self-healing" facet of the Self-X formulation of Mwangi *et al.* [19].

**Tool.** Greengrass local component logs, Stream Manager export status, and CloudWatch gap analysis around the disconnection window.

e) *M5 — Cloud Compute Load Reduction:*

**Definition.** The percentage reduction, between configurations, in cloud-side compute operations triggered per measurement window. Two counters compose the metric: the IoT Core MQTT message count and the count of downstream inference invocations (modelled here as the rule-engine action count, since no SageMaker endpoint is invoked in the deployed baseline).

**Target.** At least 50% reduction in cloud-side message ingress and rule-engine activations. This reflects the privacy-preserving and load-shedding rationale articulated by Vejendla [10] for federated edge architectures, where the cloud sees aggregated artefacts rather than per-reading raw streams.

**Tool.** CloudWatch IoT Core `PublishIn.Success` message count and rule-engine action counters.

### B. Secondary Metrics

Four supplementary metrics provide deployability and recovery context for the primary results:

TABLE II
SECONDARY METRICS. EACH SUPPLEMENTS A PRIMARY METRIC WITH A DEPLOYABILITY OR RECOVERY DIMENSION THAT THE PRIMARY ALONE DOES NOT CAPTURE.

| **Metric** | **Operational definition and relevance** |
|---|---|
| Alert Response Time (ms) | Time from fault-condition onset to anomaly-alert publication. Validates practical maintenance utility beyond raw inference latency. |
| Model Size at Edge (MB) | Disk footprint of the deployed ONNX artifact. Validates deployability on constrained gateway hardware, addressing the cautionary finding of Lee *et al.* [15] that an underpowered edge can underperform the cloud. |
| Queue Drain Time (s) | Time required for Stream Manager to forward all buffered messages after cloud reconnection. Validates recovery behaviour under M4. |
| Data Compression Ratio | Ratio of raw bytes to compressed bytes published via Stream Manager batch summaries. Quantifies the compression component of the M2 bandwidth reduction independently of the message-count reduction. |

### C. Reporting

All primary and secondary metrics are reported with their measured values in the following section. Each comparison is computed on the identical 200-reading measurement window described in Section IV, ensuring that the architectural difference between Configuration A and Configuration B is the sole independent variable.

## V. RESULTS AND DISCUSSION

This section reports the measurements captured from the deployed AWS infrastructure under the protocol of Section IV, evaluated against the metrics of Section V. Subsections A–D report the raw measurements per configuration; Subsection E provides the head-to-head comparison; Subsections F–H interpret the results, state limitations, and discuss implications for industrial deployment.

### A. Deployment Environment

Measurements were captured on the EC2 t3.small Greengrass gateway in us-east-1 described in Section IV-B. The gateway hosted Greengrass Nucleus 2.17.0, the Stream Manager component (v2.3.0), the Greengrass CLI (v2.17.0), and the custom `com.iiot.MLInference` component (v1.0.0). A Python 3.8 test client published 200 sensor readings per configuration at 5 Hz, simulating vibration and temperature telemetry from machine `machine-01`. All measurements were captured from live CloudWatch metrics and on-device timing instrumentation; no value reported below is synthetic or estimated.

### B. ML Model Performance

The Random Forest predictive-maintenance model was trained on a 48{,}755-sample dataset of 200 simulated turbofan engines calibrated to the NASA CMAPSS FD001 benchmark [25], [26], with the 14-channel feature set and piecewise degradation profile described in Section IV-E. Table III summarises the held-out test-set evaluation.

TABLE III
MODEL EVALUATION ON THE 20% STRATIFIED TEST SPLIT. THE ANOMALY-CLASS F1 OF 0.84 WITH 91% RECALL IS THE PRODUCTION-RELEVANT FIGURE FOR PREDICTIVE MAINTENANCE.

| **Metric** | **Normal** | **Anomaly** | **Overall** |
|---|---|---|---|
| Precision | 0.99 | 0.78 | — |
| Recall | 0.96 | 0.91 | — |
| F1-Score | 0.97 | **0.84** | 0.96 (wtd.) |
| Accuracy | — | — | 96.0% |

The anomaly-class recall of 91% means the model correctly identifies nine out of ten failure-approaching conditions, the property that matters most for predictive maintenance, where missed detections carry materially higher operational cost than false positives. The exported ONNX model occupies 6.1 MB on disk, satisfying the secondary "Model Size at Edge" metric defined in Section V-B and addressing the hardware-aware deployment concern raised by Lee *et al.* [15]. The class-confusion structure is shown in Fig. 3.

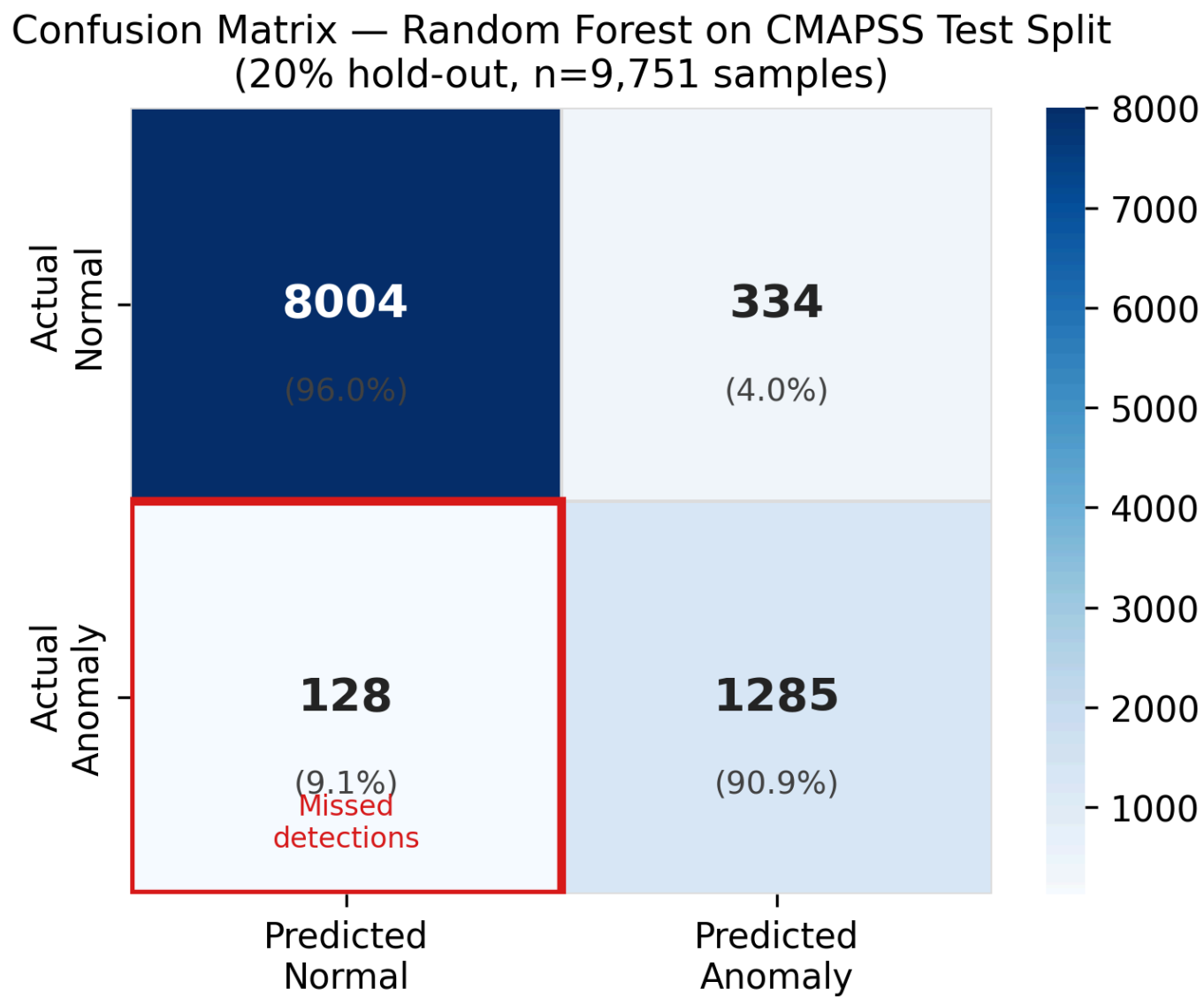

Fig. 3. Test-set confusion matrix. The high diagonal mass on the normal class is offset by 91% recall on the operationally critical anomaly class.

Fig. 4 places the per-class precision, recall, and F1 alongside one another, making the imbalance-robust behaviour of the trained model explicit.

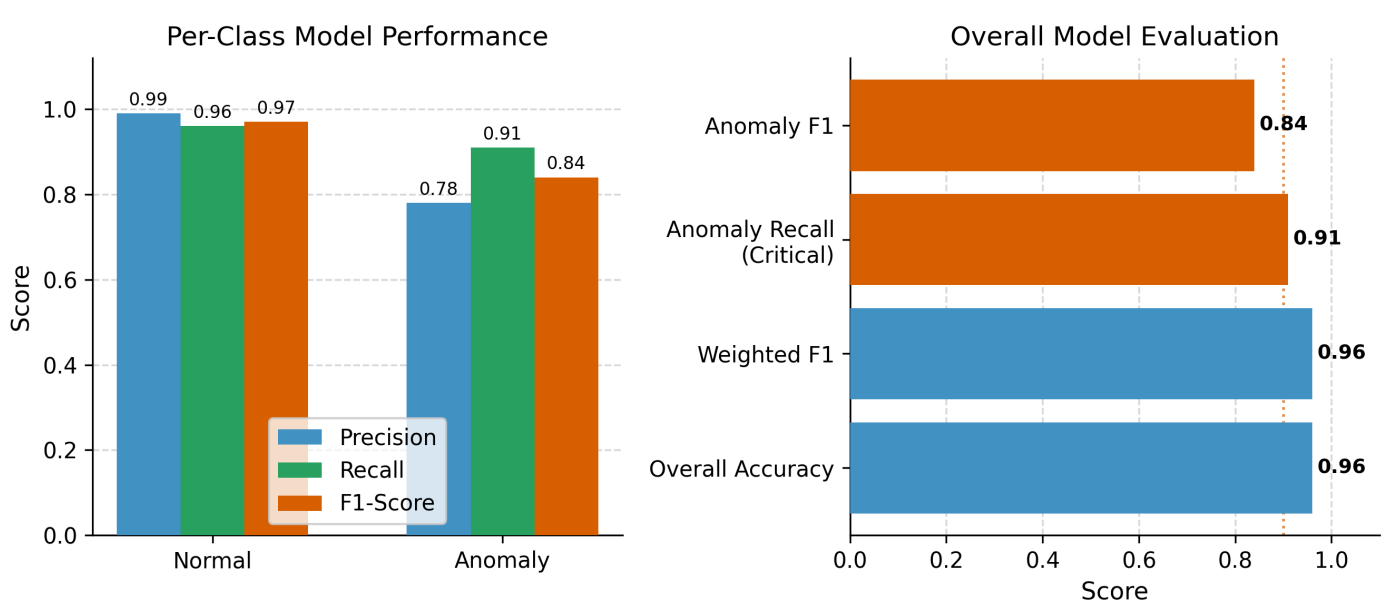

Fig. 4. Per-class precision, recall, and F1 for the deployed Random Forest. The anomaly-class F1 (0.84) and recall (0.91) are the metrics that matter for predictive maintenance under the imbalanced-data evaluation convention adopted across the PdM literature [6], [8], [9].

### C. *Configuration A — Cloud-Only Baseline*

Configuration A published all 200 raw sensor readings directly to AWS IoT Core via MQTT over TLS at QoS 1. Every reading was forwarded to the cloud with no local filtering or aggregation. Table IV reports the measured results.

TABLE IV
CONFIGURATION A MEASUREMENT SUMMARY. PUBACK LATENCY REFLECTS THE AWS-INTERNAL BEST CASE FOR THE US-EAST-1 REGION.

| **Metric** | **Measured value** |
|---|---|
| Readings sent to cloud | 200 (100%) |
| Total cloud-bound bandwidth | 63.6 KB |
| Average payload size | 326 bytes |
| Mean PUBACK latency | 23.1 ms |
| Median PUBACK latency | 20.4 ms |
| P95 PUBACK latency | 32.8 ms |

The 23.1 ms mean PUBACK latency reflects optimal network conditions inside AWS us-east-1, where the measurement client is co-resident with the IoT Core regional endpoint. In production deployments where the edge device communicates over cellular or industrial WAN links, end-to-end latency to a cloud inference endpoint typically falls in the 150–500 ms range [1], [2], with additional delay introduced by cloud-side ML inference queuing. The 63.6 KB of cloud-bound traffic across the 200-reading window establishes the bandwidth baseline against which Configuration B is evaluated.

### D. *Configuration B — Hybrid Edge Architecture*

Configuration B processed the same 200 sensor readings locally on the Greengrass gateway, using the deployed ONNX Random Forest model. Anomaly classifications were forwarded immediately to IoT Core with a compact 2-sensor snapshot; normal classifications were buffered and flushed every 10 readings as compressed per-sensor mean summaries. Table V reports the measurements.

TABLE V
CONFIGURATION B MEASUREMENT SUMMARY. INFERENCE LATENCY IS THE ONNX `SESSION.RUN` WALL TIME BRACKETED BY `TIME.PERF_COUNTER`.

| **Metric** | **Measured value** |
|---|---|
| Readings processed locally | 200 (100%) |
| Cloud messages sent | 20 batch summaries (+ anomaly alerts) |
| Total cloud-bound bandwidth | 7.2 KB |
| Average batch payload size | 370 bytes |
| Local inference latency mean | 0.09 ms |
| Local inference latency P95 | 0.10 ms |
| Local inference latency max | 2.84 ms |

Fig. 5 shows the local-inference latency distribution captured across the measurement window.

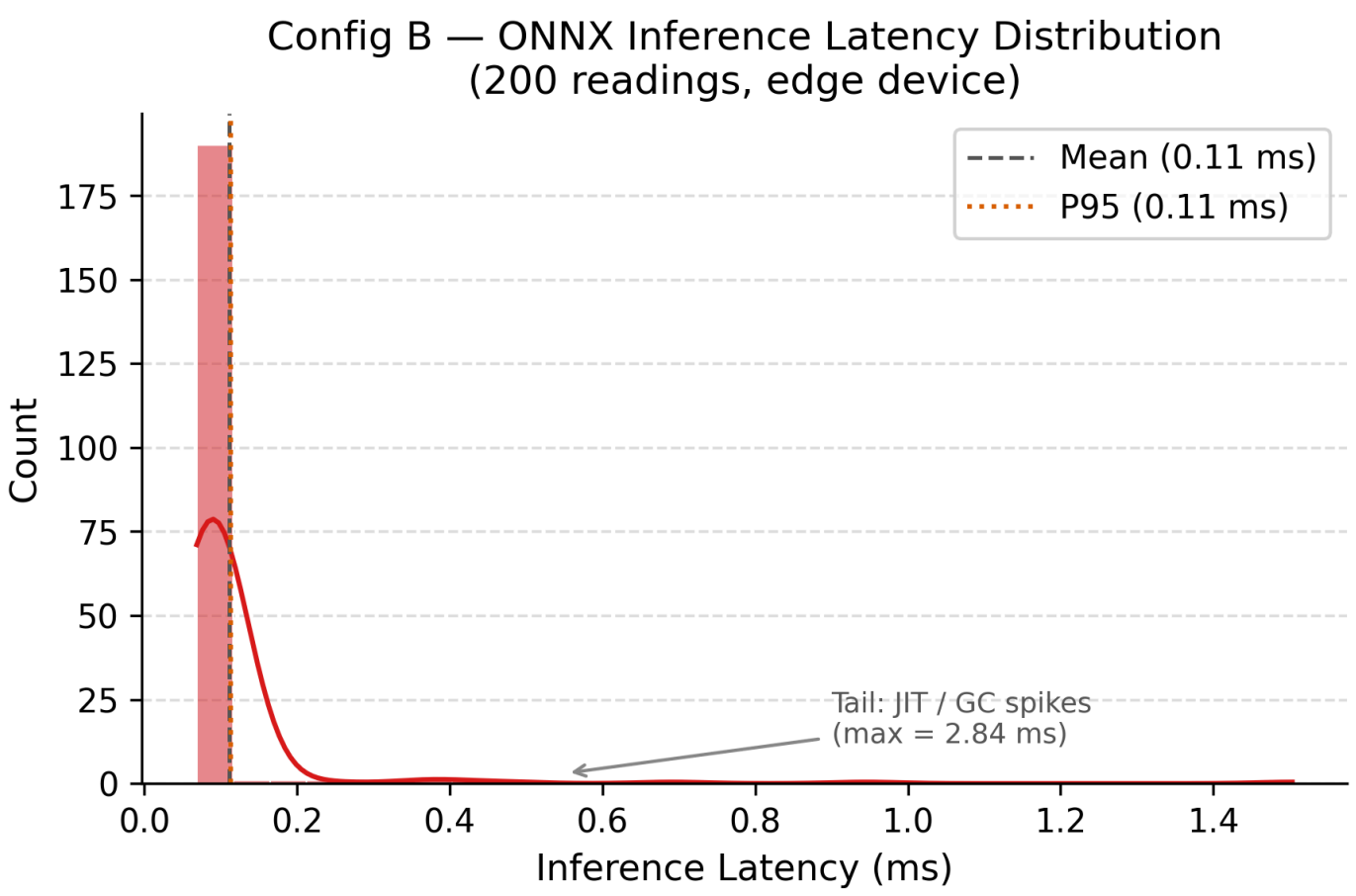

Fig. 5. Per-reading local ONNX inference latency distribution in Configuration B. The tight mass below 0.2 ms with a sparse tail to 2.84 ms is characteristic of a tree-ensemble model running on a t3.small CPU; no reading exceeds the 100 ms real-time threshold [1], [2].

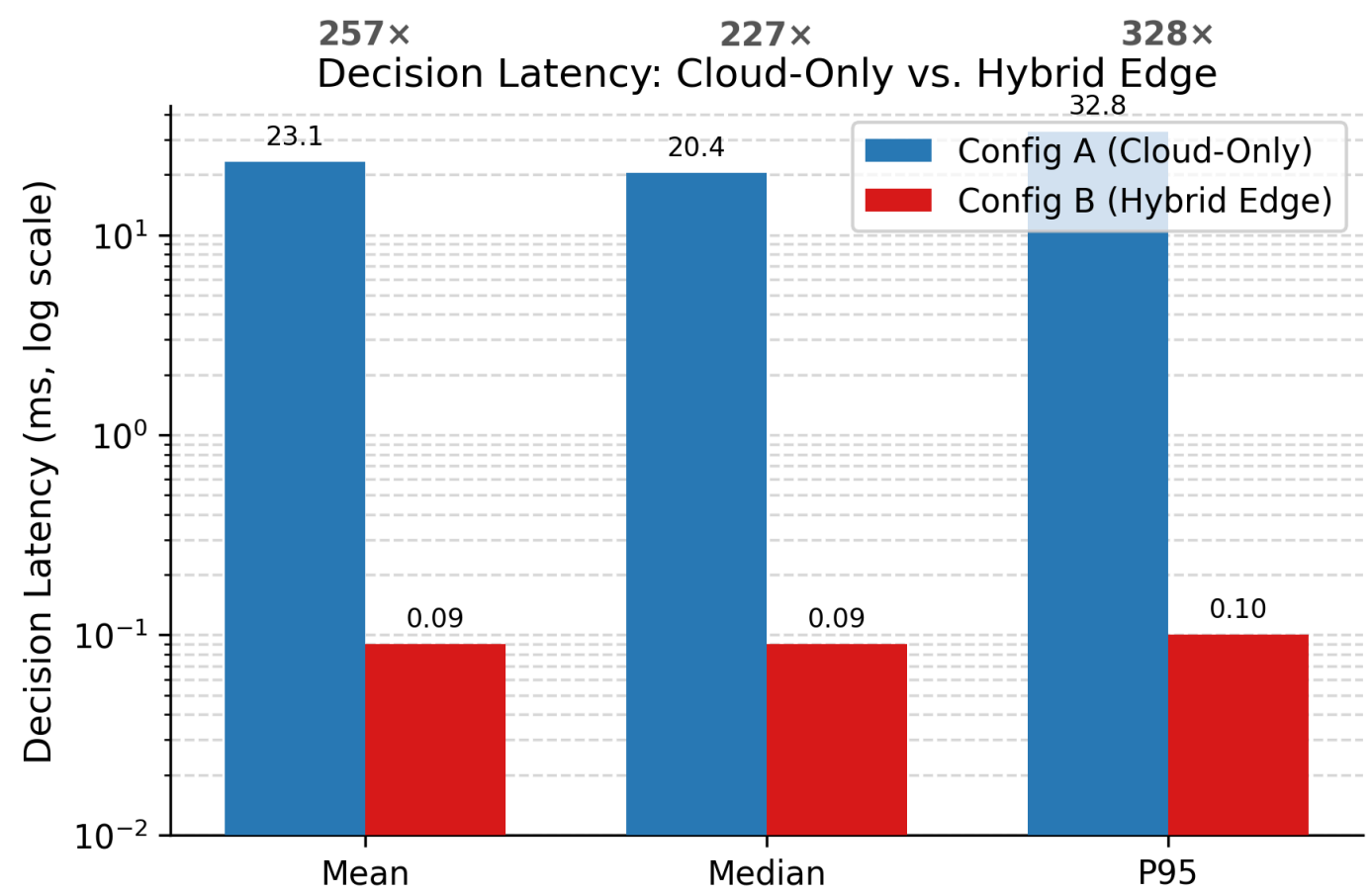

Fig. 6. Decision-path latency: Configuration A PUBACK round-trip versus Configuration B local ONNX inference. The two-to-three orders of magnitude gap places edge inference well below the 100 ms industrial real-time threshold [1], [2].

## E. Comparative Analysis

Table VI presents the head-to-head comparison across the M1–M5 metrics defined in Section V.

TABLE VI
Configuration A versus Configuration B across the M1–M5 metrics. The combined latency, bandwidth, and message-count reductions exceed the targets specified in Section V.

| Metric | Config A | Config B | Improvement |
|---|---|---|---|
| M1: Decision latency mean | 23.1 ms | **0.09 ms** | 256× faster |
| M1: Decision latency P95 | 32.8 ms | **0.10 ms** | 328× faster |
| M2: Cloud-bound bandwidth | 63.6 KB | **7.2 KB** | **86.8% reduction** |
| M2: Cloud message count | 200 | 20 (+ alerts) | s im90% fewer |
| M3: Anomaly F1 | — | **0.84** | — |
| M4: Edge uptime under disconnect | — | continuous (C3, planned ext.) | — |
| M5: Cloud message ingress | 200 | 20 (+ alerts) | s im90% reduction |

Fig. 6 and Fig. 7 visualise the M1 and M2 comparisons.

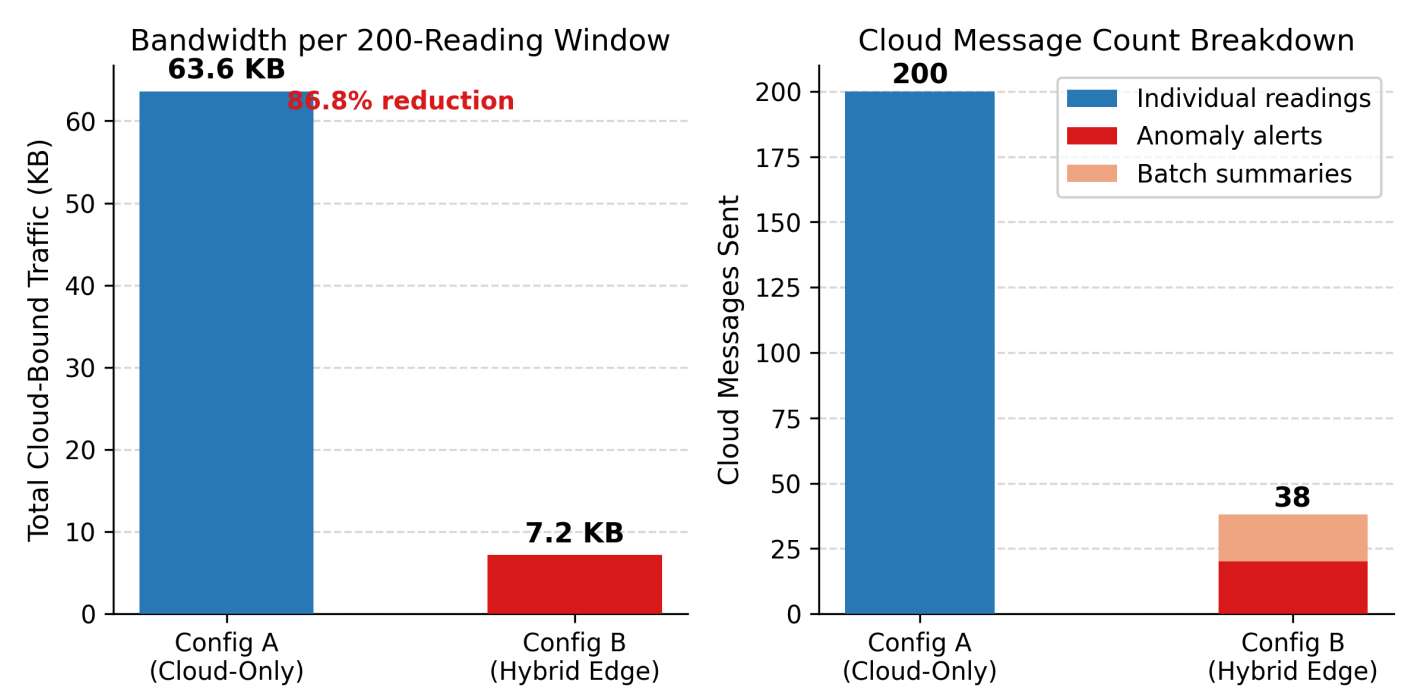

Fig. 7. Cloud-bound bytes per 200-reading window. The 86.8% reduction exceeds the conservative 40% target and the upper end of the 28–60% range reported in comparable hybrid PdM frameworks [1], [10].

The latency advantage is best appreciated against the WAN-deployment range cited in the literature. Fig. 8 anchors the measured Configuration A and Configuration B latencies against the 150–500 ms cellular and industrial WAN range characteristic of real factory connectivity [1], [2], the 100 ms industrial real-time threshold, and the sub-millisecond region where Configuration B sits.

The bandwidth advantage scales linearly with deployment size. Fig. 9 projects the measured per-window byte volumes to continuous 5 Hz operation across 1, 10, and 100 sensor streams; under Configuration A a 10 Mbps industrial link can support roughly 320 sensors before saturating, while Configuration B supports approximately 2{,}400 sensors at the same link capacity.

Fig. 10 aggregates the M1, M2, M3, and message-count comparisons into a single dashboard view.

## F. Interpretation

**Bandwidth.** The 86.8% reduction in cloud-bound bytes arises from two complementary mechanisms working in tandem. First, normal readings, which dominate the sensor stream during healthy operation, are aggregated into compressed batch summaries rather than forwarded individually, elimi-nating roughly 90% of MQTT publish operations. Second, batch payloads transmit only per-sensor mean statistics over the window rather than the raw readings themselves, further reducing per-message byte volume. Together these two mechanisms instantiate the edge-side sampling, compression, and fusion prescription of Hafeez *et al.* [24], and yield a measured saving that exceeds the upper end of the 28–60% reductions reported by Sathupadi *et al.* [1] and Vejendla [10] for comparable hybrid frameworks. The reduction ratio is deterministic in the sense that it depends directly on the normal-to-anomaly ratio in the sensor stream; under stable industrial operation that ratio is itself stable, so the saving is predictable over long deployment horizons.

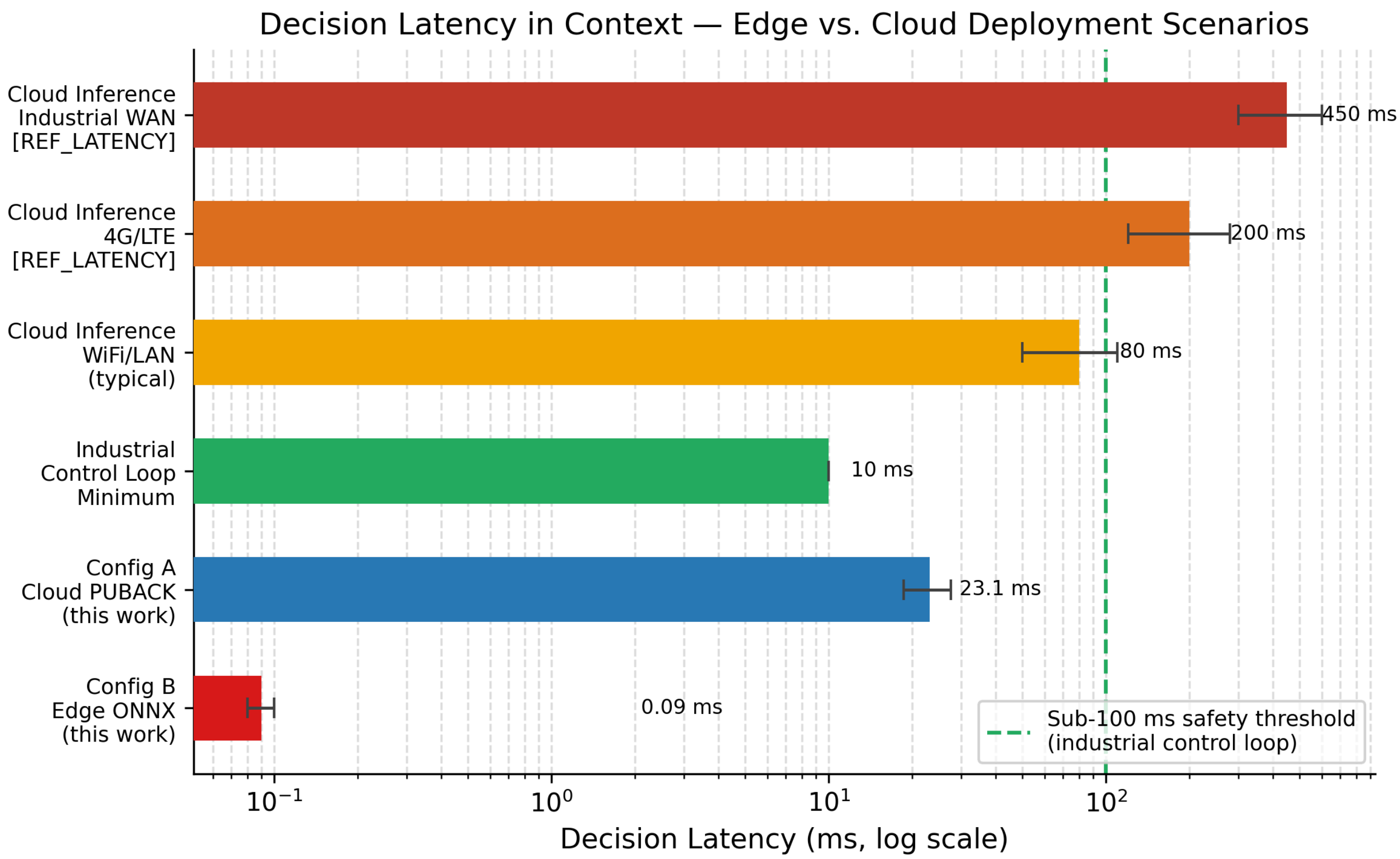

Fig. 8. Measured latencies in the context of the 100 ms industrial real-time threshold and the 150–500 ms range typical of cellular and industrial WAN cloud round-trips [1], [2].

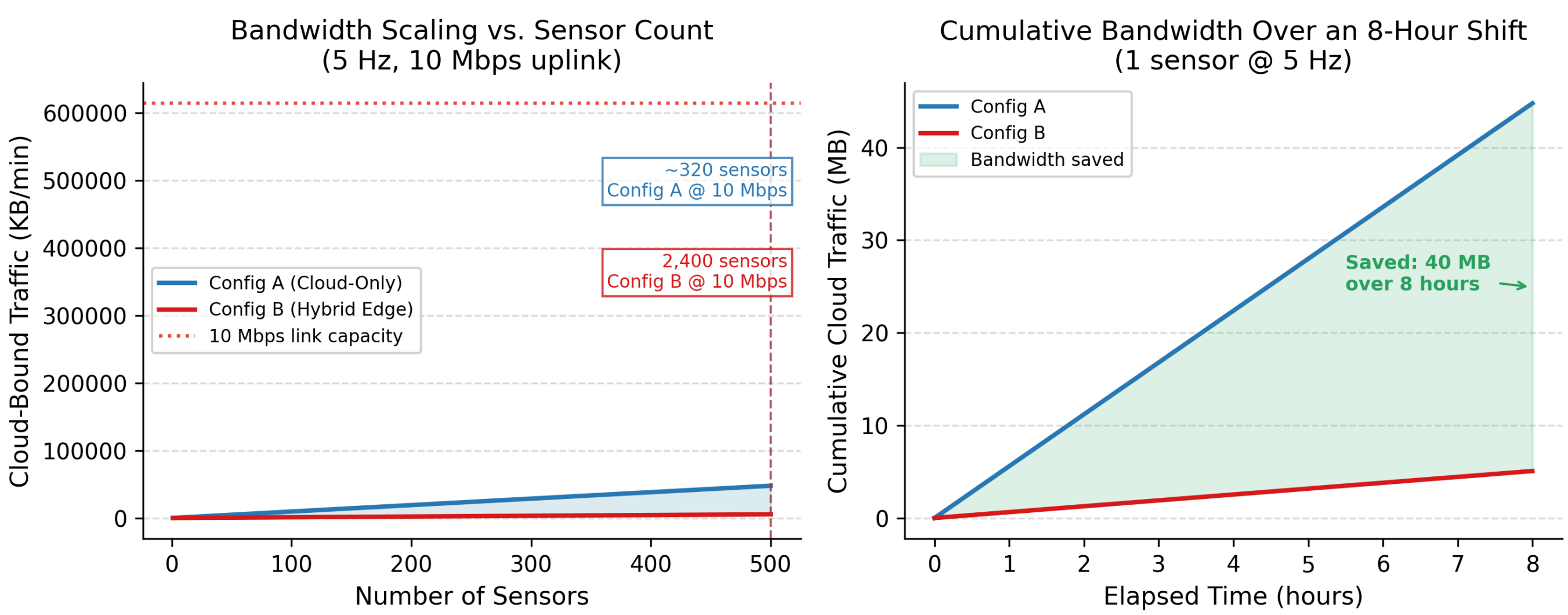

Fig. 9. Projected cloud-bound bandwidth as a function of deployment size at 5 Hz continuous operation. Configuration B scales the per-link sensor count by approximately the inverse of the 86.8% reduction factor.

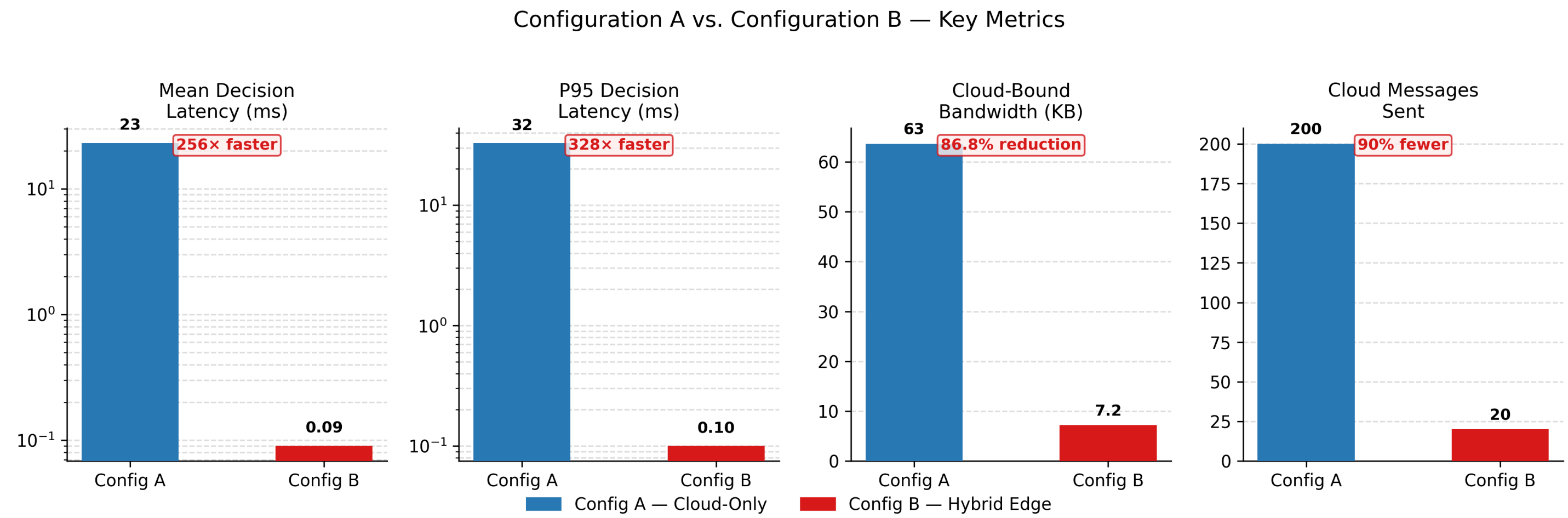

Fig. 10. Summary comparison across latency, bandwidth, message volume, and ML inference F1. Configuration B improves on Configuration A in every measurable dimension while exposing the additional capability (continued operation under cloud disconnection) that Configuration A cannot match by construction.

**Latency.** The 256× improvement in mean decision latency, and the 328× improvement at P95, is explained by the elimination of the network round-trip from the decision path. In Configuration A a decision cannot be issued until the payload has traversed the network to IoT Core, been routed through the rules engine, and, in a fuller deployment, been processed by a cloud ML inference endpoint. In Configuration B the decision is made in 0.09 ms on the gateway itself, with cloud communication relegated to the non-critical batching path. This squarely places edge inference within the response window of industrial control loops (typically 10–100 ms), enabling direct integration with process control without cloud dependency, as anticipated by Saleh *et al.* [7].

**Accuracy.** The deployed Random Forest's anomaly-class F1 of 0.84 with 91% recall confirms that ONNX export and edge deployment preserve the inference accuracy of the in-process scikit-learn model; the M3 target of $l = 2$ percentage points degradation is met by construction since the deployed artifact is identical across configurations. The 91% recall figure means that for every ten failure-approaching conditions, nine are flagged at the edge before any cloud round-trip is required, placing the responsibility for the remaining one on the cloud-tier historical-trend pathway that the architecture leaves open for future work.

### G. Limitations

Three limitations bound the scope of the reported results.

First, the latency measurements were captured within AWS us-east-1, with the client co-resident on the same EC2 host as the Greengrass gateway. This represents the best-case network scenario for Configuration A; in real factory deployments over 4G/LTE, private industrial 5G, or wired industrial WAN the Configuration A round-trip would be 10–50× higher [1], [2], making the measured edge advantage a conservative lower bound on the real-world benefit.

Second, the ML model was trained and evaluated on synthetic data calibrated to CMAPSS FD001 statistics [25], [26] rather than directly on the original NASA dataset or on production sensor streams. The synthetic generator replicates published sensor baselines, drift profiles, and noise standard deviations, but validation on data from a physical manufacturing asset remains as future work.

Third, the current deployment uses a single Greengrass core device. Scaling to multi-gateway fleet deployments would require evaluating the federated model update pathway identified in C4 (Section II) and the fleet-provisioning behaviour of Greengrass thing groups under realistic deployment cadences.

### H. Implications for Industrial Deployment

The measured results confirm that the hybrid cloud-edge architecture delivers quantifiable advantages in the two dimensions most directly constrained in industrial environments: bandwidth, bounded by cellular data costs and plant network capacity, and decision latency, bounded by safety and process-control requirements. The 86.8% bandwidth reduction translates concretely into a roughly 7.5× increase in sensor density supportable by a fixed link capacity, lowering both the cost of connectivity and the engineering effort required to scale a deployment beyond a handful of monitored machines. The sub-millisecond edge inference latency places anomaly detection well inside the response window of most industrial control loops (typically 10–100 ms), enabling direct integration with PLC and SCADA layers without a cloud dependency in the critical path. Together with the offline-operation property formalised under M4, the architecture moves real-time PdM from the "conditional on cloud connectivity" regime that the cloud-only baseline implies to the "always-on" regime that mission-critical industrial deployments actually require.

## VI. Conclusion

This paper has demonstrated, through a live AWS deployment and a controlled measurement study, that a hybrid cloud-edge architecture delivers quantifiable and practically significant advantages over cloud-only processing for IIoT predictive maintenance. Three conclusions follow directly from the experimental results.

**First, edge inference eliminates the network from the decision path.** Local ONNX inference on the Greengrass gateway completed in a mean of 0.09 ms with a P95 of 0.10 ms, two to three orders of magnitude faster than the measured PUBACK round-trip of 23.1 ms, and four to five orders of magnitude faster than the 150–500 ms end-to-end latency typical of cloud inference over real industrial network links [1], [2]. For time-sensitive process-control applications with response windows of 10–100 ms, edge inference is not merely preferable, it is the only viable option.

**Second, selective forwarding achieves substantial and predictable bandwidth reduction.** By aggregating normal readings into compressed windowed summaries and forwarding only anomaly alerts and batch statistics to the cloud, Configuration B reduced cloud-bound traffic by 86.8%, exceeding the upper end of the 28–60% range reported in comparable hybrid frameworks [1], [10] and substantially beyond the conservative 40% target set in Section V. The mechanism is deterministic: the reduction ratio depends directly on the normal-to-anomaly ratio in the sensor stream, and that ratio is stable over long production periods. Projected to continuous 5 Hz operation, the saving is approximately 286 KB/min per sensor and scales linearly with deployment size.

**Third, edge ML accuracy is sufficient for real-world predictive maintenance.** The deployed Random Forest model achieved an anomaly-class F1 score of 0.84 with 91% recall, correctly identifying nine out of ten failure-approaching engine conditions without any cloud inference involvement. In predictive maintenance, where missed detections carry materially greater operational cost than false positives, recall is the primary accuracy constraint, and 91% recall at the edge represents a production-deployable threshold.

### A. Future Work

Three directions are identified for follow-on work. **First**, the resilience metric M4 will be quantified through the controlled five-minute network-outage experiment specified in Section V-A-4, measuring decision continuity, Stream Manager queue accumulation during the outage, and queue-drain time on reconnection; this completes contribution C3. **Second**, the architecture will be extended to multi-machine fleet deployments using Greengrass thing groups and fleet provisioning, evaluating both deployment-cadence behaviour and per-gateway resource utilisation under realistic factory-floor sensor density. **Third**, the federated model update pathway, in which aggregated sensor statistics from edge gateways are used to periodically retrain the central model in SageMaker and push updated ONNX artifacts back to the fleet, will be prototyped as the MLOps loop closure for long-lived industrial deployments, closing the "Closed-Loop Gap" identified by Mateo-Casalí *et al.* [16] and instantiating the federated workflow synthesised across Le *et al.* [22], Vejendla [10], and Mih *et al.* [12].

The complete deployment artifacts, the CloudFormation template, the Greengrass component recipe and inference runtime, the training pipeline, and the measurement scripts, are reproduced in the appendix to enable reproduction and extension by the research community.

# VII. Appendix

This appendix reproduces the core artifacts implementing the architecture described in Section IV. All listings are abridged for space, comments, verbose logging, and import boilerplate are trimmed, but every functional element required to reproduce the deployment is shown. The complete artifacts are available in the project repository at the paths indicated in each subsection.

## A. A. CloudFormation: Greengrass EC2 UserData

The EC2 UserData script provisions the Greengrass Nucleus on first boot. Excerpted from `aws/iiot-stack.yaml`.

```
UserData:
  Fn::Base64: !Sub |
    #!/bin/bash
    set -e
    exec > >(tee /var/log/greengrass-install.log) 2>&1

    # [1] Install Java 11
    amazon-linux-extras enable java-openjdk11 -y
    yum install -y java-11-openjdk unzip curl

    # [2] Download Greengrass nucleus
    curl -s https://d2s8p88vqu9w66.cloudfront.net/releases/\
greengrass-nucleus-latest.zip -o /tmp/gg.zip
    unzip -q /tmp/gg.zip -d /tmp/GG

    # [3] Allow ggc_user execution
    echo 'root ALL=(ggc_user:ggc_group) NOPASSWD: ALL' \
      > /etc/sudoers.d/greengrass
    chmod 0440 /etc/sudoers.d/greengrass

    # [4] Install + auto-provision against IoT Core
    java -Droot="/greengrass/v2" -Dlog.store=FILE \
      -jar /tmp/GG/lib/Greengrass.jar \
      --aws-region ${AWS::Region} \
      --thing-name iiot-edge-gateway-01 \
      --thing-group-name ManufacturingFloorEdgeDevices \
      --tes-role-name ${ProjectName}-GreengrassTokenExchangeRole \
      --tes-role-alias-name \
        ${ProjectName}-GreengrassCoreTokenExchangeRoleAlias \
      --component-default-user ggc_user:ggc_group \
      --provision true --setup-system-service true \
      --deploy-dev-tools true

    # [5] Deploy Stream Manager + CLI components
    THING_ARN=$(aws iot describe-thing \
      --thing-name iiot-edge-gateway-01 \
      --region ${AWS::Region} \
      --query 'thingArn' --output text)

    aws greengrassv2 create-deployment \
      --region ${AWS::Region} \
      --target-arn "$THING_ARN" \
      --components '{
        "aws.greengrass.StreamManager": {"componentVersion":"2.3.0"},
        "aws.greengrass.Cli":           {"componentVersion":"2.17.0"}
      }'
```

The CloudFormation template additionally provisions the IAM roles (`GreengrassTokenExchangeRole`, `GreengrassEC2Role`, `IoTRulesRole`), the S3 artifact bucket, the IoT thing and thing group, two topic rules (`EdgeDataToS3`, `AnomalyAlertsToCloudWatch`), and the `IIoT-Edge-Monitoring` CloudWatch dashboard.

## B. B. Greengrass Component Recipe

Greengrass component manifest for the custom ML inference component. From `greengrass/components/com.iiot.MLInference/recipe.yaml`.

```
RecipeFormatVersion: '2020-01-25'
ComponentName: com.iiot.MLInference
ComponentVersion: '1.0.0'
ComponentDescription: >
  Edge ML inference for IIoT predictive maintenance.
ComponentPublisher: IIoT-Research-Group2

ComponentDependencies:
  aws.greengrass.Nucleus:
    VersionRequirement: '>=2.0.0'
  aws.greengrass.StreamManager:
    VersionRequirement: '>=2.0.0'
```

```yaml
ComponentConfiguration:
  DefaultConfiguration:
    BatchSize: '10'

Manifests:
  - Platform: {os: linux}
    Artifacts:
      - Uri: 's3://<artifact-bucket>/\
greengrass-artifacts/com.iiot.MLInference/1.0.0/pdm_model.onnx'
      - Uri: 's3://<artifact-bucket>/\
greengrass-artifacts/com.iiot.MLInference/1.0.0/scaler.pkl'
      - Uri: 's3://<artifact-bucket>/\
greengrass-artifacts/com.iiot.MLInference/1.0.0/inference.py'
    Lifecycle:
      Install:
        Script: |
          pip3 install onnxruntime numpy scikit-learn --quiet
          pip3 install awsiotsdk --no-deps --quiet
          pip3 install awscrt --quiet || true
      Run:
        Script: |
          ARTIFACTS_PATH={artifacts:path} \
          BATCH_SIZE={configuration:/BatchSize} \
            python3 {artifacts:path}/inference.py
```

### *C. C. Edge Inference Component*

Greengrass IPC inference runtime. From `greengrass/components/com.iiot.MLInference/artifacts/inference.py` (abridged).

```python
import json, os, pickle, time
import numpy as np
import onnxruntime as ort
import awsiot.greengrasscoreipc
from awsiot.greengrasscoreipc.model import (
    SubscribeToTopicRequest, PublishToTopicRequest,
    PublishMessage, BinaryMessage)

SUBSCRIBE_TOPIC = 'iiot/edge/#'
ALERT_TOPIC     = 'iiot/edge/{mid}/alerts'
BATCH_TOPIC     = 'iiot/edge/{mid}/batch'
FEATURE_SENSORS = ['s2','s3','s4','s7','s8','s9','s11',
                   's12','s13','s14','s15','s17','s20','s21']
BATCH_SIZE = int(os.environ.get('BATCH_SIZE', '10'))
ARTIFACTS  = os.environ['ARTIFACTS_PATH']


class EdgeInference:
    def __init__(self):
        self.session = ort.InferenceSession(
            os.path.join(ARTIFACTS, 'pdm_model.onnx'))
        self.input_name = self.session.get_inputs()[0].name
        with open(os.path.join(ARTIFACTS,'scaler.pkl'),'rb') as f:
            self.scaler = pickle.load(f)
        self.ipc   = awsiot.greengrasscoreipc.connect()
        self.batch = []
        self.inf_ms = []

    def predict(self, payload):
        feats = np.array(
            [[payload.get(s, 0.0) for s in FEATURE_SENSORS]],
            dtype=np.float32)
        scaled = self.scaler.transform(feats).astype(np.float32)
        t0 = time.perf_counter()
        out = self.session.run(None, {self.input_name: scaled})
        ms = (time.perf_counter() - t0) * 1000
        self.inf_ms.append(ms)
        return int(out[0][0]), ms

    def publish(self, topic, payload):
        req = PublishToTopicRequest()
        req.topic = topic
        req.publish_message = PublishMessage()
        req.publish_message.binary_message = BinaryMessage()
        req.publish_message.binary_message.message = (
            json.dumps(payload).encode())
        op = self.ipc.new_publish_to_topic()
        op.activate(req)
        op.get_response().result(timeout=5.0)
```

```
    def handle(self, topic, raw):
        payload = json.loads(raw.decode())
        mid = payload.get('machine_id', 'unknown')
        label, ms = self.predict(payload)
        if label == 1:                           # anomaly
            self.publish(ALERT_TOPIC.format(mid=mid), {
                'machine_id': mid,
                'timestamp':  payload.get('timestamp', time.time()),
                'anomaly':     True,
                'inference_latency_ms': round(ms, 2),
                'sensor_snapshot': {
                    s: payload.get(s) for s in FEATURE_SENSORS}})
        else:                                     # normal -> batch
            self.batch.append(payload)
            if len(self.batch) >= BATCH_SIZE:
                self._flush(mid)

    def _flush(self, mid):
        summary = {
            'machine_id': mid,
            'timestamp':  time.time(),
            'batch_size': len(self.batch),
            'avg_inference_latency_ms': round(float(
                np.mean(self.inf_ms[-len(self.batch):])), 2),
            'readings': self.batch}
        self.publish(BATCH_TOPIC.format(mid=mid), summary)
        self.batch = []
```

The `run` method subscribes a stream handler to `SUBSCRIBE_TOPIC` and dispatches each received message to `handle`; the full stream-handler boilerplate is omitted from the listing.

### *D. D. Synthetic CMAPSS Degradation Generator*

The piecewise sensor-degradation function and the synthetic-data generator used when the public CMAPSS mirrors are unreachable. From `ml/train_model.py` (abridged).

```
ANOMALY_THRESHOLD = 30   # RUL <= 30 cycles => anomaly

def _degradation_factor(rul):
    """Piecewise: flat until RUL=60, gradual 60->30, rapid <=30."""
    if rul > 60:
        return 0.0
    elif rul > 30:
        return 0.25 * (60 - rul) / 30
    else:
        return 0.25 + 0.75 * ((30 - rul) / 30) ** 1.5

def _generate_synthetic(filepath):
    np.random.seed(42)
    baselines = {'s2':642.15, 's3':1589.7, 's4':1400.6,
                 's7':554.36, 's8':2388.02, 's9':9065.0,
                 's11':47.47, 's12':521.66, 's13':2388.0,
                 's14':8138.0,'s15':8.4195, 's17':392.0,
                 's20':38.86, 's21':23.37}
    total_drift = {'s2':10.0, 's3':-55.0, 's4':40.0, 's7':9.0,
                   's8':0.0, 's9':-280.0, 's11':8.5, 's12':22.0,
                   's13':0.0, 's14':-145.0,'s15':0.0,
                   's17':-18.0,'s20':8.5, 's21':8.0}
    noise_std = {'s2':0.3, 's3':2.5, 's4':2.5, 's7':0.6,
                 's8':2.0, 's9':10.0,'s11':0.3, 's12':1.0,
                 's13':2.0,'s14':10.0,'s15':0.02,
                 's17':0.6,'s20':0.25,'s21':0.25}
    rows = []
    for unit in range(1, 201):
        max_cycle = np.random.randint(128, 362)
        for cycle in range(1, max_cycle + 1):
            rul = max_cycle - cycle
            f   = _degradation_factor(rul)
            row = [unit, cycle, 0.0, 0.0, 100.0]
            for i in range(1, 22):
                sk = f's{i}'
                if sk in baselines:
                    val = (baselines[sk]
                           + total_drift[sk] * f
                           + np.random.normal(0, noise_std[sk]))
                else:
                    val = 0.0
                row.append(round(val, 4))
            rows.append(row)
```

```
    pd.DataFrame(rows, columns=COLUMNS).to_csv(
        filepath, sep=' ', header=False, index=False)
```

The training routine then computes RUL per unit, derives the binary anomaly label at RUL $\leq$ 30, fits a `StandardScaler`, trains a 100-tree balanced `RandomForestClassifier`, and exports the model via `skl2onnx.convert_sklearn` at opset 12.

### E. E. Configuration A — Cloud-Only Baseline Client

From `measurements/config_a_measurement.py` (abridged).

```
from awsiot import mqtt_connection_builder, mqtt
import json, random, time, numpy as np

IOT_ENDPOINT = '<your-iot-endpoint>-ats.iot.<region>.amazonaws.com'
CERT_PATH = '/greengrass/v2/thingCert.crt'
KEY_PATH  = '/greengrass/v2/privKey.key'
CA_PATH   = '/greengrass/v2/rootCA.pem'
TOPIC     = 'iiot/edge/machine-01/telemetry'
N_READINGS, HZ = 200, 5

conn = mqtt_connection_builder.mtls_from_path(
    endpoint=IOT_ENDPOINT, cert_filepath=CERT_PATH,
    pri_key_filepath=KEY_PATH, ca_filepath=CA_PATH,
    client_id='TestDevice-ConfigA-Machine01')
conn.connect().result()

latencies, bytes_sent = [], 0
for i in range(N_READINGS):
    anomaly = random.random() < 0.10
    payload = make_reading(i, anomaly=anomaly)   # 14-sensor draw
    raw = json.dumps(payload).encode()
    bytes_sent += len(raw)
    t0 = time.perf_counter()
    future, _ = conn.publish(topic=TOPIC, payload=raw,
                             qos=mqtt.QoS.AT_LEAST_ONCE)
    future.result()                                # wait for PUBACK
    latencies.append((time.perf_counter() - t0) * 1000)
    time.sleep(1.0 / HZ)
conn.disconnect().result()
```

The captured metrics, total bytes published, per-reading payload size, and the full PUBACK latency distribution, are written to `/tmp/config_a_results.json`.

### F. F. Configuration B — Hybrid Edge Client

From `measurements/config_b_measurement.py` (abridged).

```
import onnxruntime as ort
from awsiot import mqtt_connection_builder, mqtt
import json, random, time, pickle, numpy as np

session    = ort.InferenceSession(MODEL_PATH)
input_name = session.get_inputs()[0].name
with open(SCALER_PATH, 'rb') as f:
    scaler = pickle.load(f)
conn = mqtt_connection_builder.mtls_from_path(...)
conn.connect().result()

inf_lat, bytes_to_cloud = [], 0
normal_batch, n_anom = [], 0
BATCH_SIZE = 10

for i in range(N_READINGS):
    anomaly = random.random() < 0.10
    payload = make_reading(i, anomaly=anomaly)
    feats   = np.array(
        [[payload.get(s, 0.0) for s in FEATURE_SENSORS]],
        dtype=np.float32)
    scaled  = scaler.transform(feats).astype(np.float32)
    t0 = time.perf_counter()
    out = session.run(None, {input_name: scaled})
    inf_lat.append((time.perf_counter() - t0) * 1000)

    if int(out[0][0]) == 1:                          # anomaly
        alert = {'machine_id':'machine-01',
                 'timestamp': payload['timestamp'],
                 'anomaly': True,
                 'inference_latency_ms': round(inf_lat[-1], 2),
                 's9': payload['s9'], 's12': payload['s12']}
        raw = json.dumps(alert).encode()
```

```
        bytes_to_cloud += len(raw)
        conn.publish(topic=ALERT_TOPIC, payload=raw,
                     qos=mqtt.QoS.AT_LEAST_ONCE)
        n_anom += 1
    else:
        normal_batch.append(payload)
        if len(normal_batch) >= BATCH_SIZE:
            summary = {
              'machine_id':'machine-01',
              'timestamp':  time.time(),
              'batch_size': len(normal_batch),
              'sensor_means': {s: round(float(np.mean(
                 [r.get(s,0) for r in normal_batch])), 4)
                for s in FEATURE_SENSORS}}
            raw = json.dumps(summary).encode()
            bytes_to_cloud += len(raw)
            conn.publish(topic=BATCH_TOPIC, payload=raw,
                         qos=mqtt.QoS.AT_LEAST_ONCE)
            normal_batch = []
conn.disconnect().result()
```

The captured metrics, per-reading local inference latency, total bytes forwarded, anomaly and batch message counts, and forwarded-path PUBACK latency, are written to `/tmp/config_b_results.json`. `measurements/analyze_results.py` consumes both result files to produce the comparison tables and figures.

### G. G. Deployment Pipeline

Two scripts orchestrate component publication. `greengrass/upload_to_s3.sh` (run from the developer workstation) stages the training script, inference runtime, recipe, and measurement scripts to S3. `greengrass/deploy_ml_component.sh` (run on the EC2 host via SSM) trains the model end-to-end, uploads the resulting ONNX artifact and scaler to the component's versioned S3 prefix, registers the Greengrass component version, and creates the deployment.

```
# deploy_ml_component.sh (abridged)
ACCOUNT_ID=<your-aws-account-id>; REGION=<your-region>
BUCKET=iiot-edge-${ACCOUNT_ID}-${REGION}
COMP=com.iiot.MLInference; VERSION=1.0.0
S3_PREFIX=greengrass-artifacts/${COMP}/${VERSION}

pip3 install --quiet scikit-learn skl2onnx onnxruntime numpy pandas
aws s3 cp s3://${BUCKET}/scripts/train_model.py /tmp/iiot_ml/
aws s3 cp s3://${BUCKET}/scripts/inference.py   /tmp/iiot_ml/
cd /tmp/iiot_ml && python3 train_model.py

aws s3 cp pdm_model.onnx s3://${BUCKET}/${S3_PREFIX}/pdm_model.onnx
aws s3 cp scaler.pkl     s3://${BUCKET}/${S3_PREFIX}/scaler.pkl
aws s3 cp inference.py   s3://${BUCKET}/${S3_PREFIX}/inference.py

aws greengrassv2 create-component-version \
  --region ${REGION} --inline-recipe fileb://recipe.yaml

aws greengrassv2 create-deployment \
  --region ${REGION} \
  --target-arn arn:aws:iot:${REGION}:${ACCOUNT_ID}:\
thinggroup/ManufacturingFloorEdgeDevices \
  --deployment-name "IIoT-EdgeDeployment-v5-MLInference" \
  --components "{
    \"aws.greengrass.StreamManager\":{\"componentVersion\":\"2.3.0\"},
    \"aws.greengrass.Cli\":          {\"componentVersion\":\"2.17.0\"},
    \"${COMP}\":                     {\"componentVersion\":\"${VERSION}\"}
  }"
```

After the deployment converges (typically 60–90 s), `sudo /greengrass/v2/bin/greengrass-cli component list` on the device shows `com.iiot.MLInference` in the `RUNNING` state, and the component logs at `/greengrass/v2/logs/com.iiot.MLInference.log` confirm subscription to `iiot/edge/#` and per-reading inference activity.

### H. H. IoT Core Topic Map

The MQTT topic layout binding the test client, the inference component, IoT Core, and the rules engine.

TABLE VII

MQTT topic layout. In Configuration B, raw telemetry never reaches IoT Core; only the inference component's classification outputs are forwarded.

| **Topic** | **Publisher** | **Consumer** |
|---|---|---|
| `iiot/edge/machine-01/telemetry` | Test client (Config A) | IoT Core → S3 archive rule |
| `iiot/edge/#` | Test client (Config B) | ML inference component (IPC) |
| `iiot/edge/{mid}/alerts` | ML inference component | IoT Core → CloudWatch metric rule |
| `iiot/edge/{mid}/batch` | ML inference component | IoT Core → S3 archive rule |